\documentclass{article}
\usepackage{amsfonts}
\newtheorem{theorem}{Theorem}[section]
\newtheorem{pro}{Proposition}[section]
\newtheorem{lemma}{Lemma}[section]
\newtheorem{definition}{Definition}[section]
\newtheorem{remark}{Remark}[section]

\newcommand{\QED}{\begin{flushright} QED\end{flushright}}

\title{ Compactification of the moduli space in symplectization
and hidden symmetries of its boundary}
\author{Gang Liu}

\begin{document}
\maketitle

\section{Introduction}

The purpose of this paper and the forth coming [L1], [L3] is to lay down
a foundation for a sequence of papers concerning  the moduli space
of connecting
pseudo-holomorphic
maps in the symplectization of a compact contact manifold and their
applications. In this paper, we will 
establish the comapctification of the moduli space 
of the pseudo-holomorphic maps in the 
symplectization  and 
exhibit some new  phenomenon
concerning bubbling and  the "hidden" symmetries of the boundary of 
 the comapctification. Combining with the index formula, which will be 
proved in [L3], we will show in [L3] that
 the virtual co-dimension of
the boundary  components  of the moduli space  with at least one bubble 
 is at least two, while the virtual co-dimension of
the boundary  components  of broken connecting maps of two elements
is one. In [L1], we will show that these virtual co-dimensions can be realized
in the corresponding virtual moduli cycles.
In a sequence of forth coming papers, we will give some of 
possible applications. In particular, we will define various versions of 
 index
homology for a contact manifold,  relative index homology for a symplectic
manifold with contact type boundary,
  as well as their multiplicative structures in these holmologies.
These multiplicative structures can be thought as analogies of the usual
quantum product and pants product in quantum cohomology and Floer
 cohomology.  We will also investigate the implication of these homologies
to Weinstein conjecture.

It is well-known that a family of pseudo-holomorphic maps in the 
symplectization of a compact contact manifold may develop bubbles.
Since in the symplectization the symplectic form is exact, each top bubble
necessarily has non-removable singularity at infinity, and along the end at 
infinity,
the bubble is convergent to some closed orbit of the Reeb field of the contact 
manifold. This makes the behavior of the boundary  components of 
the compactification of the moduli space
 here  very much look like the one of the broken connecting orbits
 in the usual
Floer homology. In particular, it is  believed  that the co-dimension of
 the boundary components even coming from bubbling  should be one in general.
 We will show in this paper and [L3] that in the case of
the moduli space the pseudo-holomorphic maps
 connecting at least two closed orbits at the two ends of the
 symplectization,at
least virtually,  this  belief is not true. 
%In a forth coming paper
%[L1], we will develop the virtual moduli cycle technique in this context
%and prove that  this virtual co-dimension  can be made to be the real
%co-dimension  for the associated virtual moduli cycle.
%Os.
 
Our starting point is the the following
new phenomenon concerning the bubbling of connecting pseudo-holomorphic 
maps. 
Observe that each time when a family of 
pseudo-holomorphic
maps connecting two closed orbits  splits
into a family of broken connecting maps or develops a bubble,
 there is not only
a splitting of the domain but also a
splitting of the target at same time. Therefore the
 ${\bf R}$-symmetry of the target splits into a two-dimensional or
multi-dimensional symmetries during the bubbling or splitting.
Moreover, the rates of these two types of degeneration of the domain and target
are independent to each other in general. In fact, the maximum principal
implies that in the simplest case when  such a 
family of connecting maps develops only one bubble, the image of the bubble
lies on a new component on the "left" of the original one, and 
there is also a new principal component on the left of the original 
 principal component. Note that the new "left" principal componet may be just
a trivial connecting map. However, the limit map itself is still stable.
This last kind of degeneration plays a rather special role. 
Therefore, unlike the usual  Gromov-Floer theory in 
symplectic case,  
 the bubbling here, 
splits the domain  into three components and  the target into two. Note that
this phenomenon can only happen when the pseudo-holomorphic maps involved
connect at least two closed orbits lying on the two ends of the
 symplectization.

Now using the fact that both symmetry groups  
of  
a connecting map  and     
a bubble with non-removable singularity are three dimensional,
it is easy to see that in term of the dimensions of symmetries,
bubbling has co-dimension three, while the  splitting of connecting
maps into broken ones is co-dimension two. This seems to suggest a 
rather different picture on the boundary behavior of the moduli space
in the symplectization, which is not only disprove what was believed
before but also bring us to a situation of dilemma. Namely, the situation
here is even better than the one in the usual Gromov-Floer theory
in symplectic  case. 

One the the main purpose of this paper and [L1], [L3] is to 
resolve this dilemma. In this paper, we will give some key ingradients
of the solution of the dilemma. The main body of this paper is devoted to
 to define the notion of stable maps in the symplectization
and  to use them to
establish the compactness of the moduli space of such maps. 
It turns out situation here is different from the usual
Gromov-Floer theory. There are various new phenomenons, which
have to be put into consideration in order to 
 to formulate the notion
of stable maps and various related notions.

In symplectic geometry,  one of the key ingredients to prove the compactness
of the moduli space of stable maps is the bubbling process.
It consists of three parts: the Uhlenbeck-Sacks rescaling scheme, removable
singularity lemma and the analysis concerning the behavior of the "connecting
tubes". In the case of
the symplectization of a contact manifold, we have mentioned above
that there is a new phenomenon in the bubbling process. However, as far as
the proof goes, there are still the correspondoing three parts there.
 The first  and most important  part of the
bubbling was established by
Hofer in [H]. He discovered the phenomenon of bubbling in the
 symplecitization
with bubble with non-removable singularity.
 Since the top bubble in the symplectization
always has non-removable singularity, the  corresponding second part 
of the bubbling here is about  the asymptotic
behavior of a bubble approaching to its non-removable singularities. In 
particular, it is important to know that along the end, a bubble
with non-removable singularity approaches to some closed orbit with
an exponential decay rate.  In
the case that the contact manifold is three dimensional, the desired
exponential decay estimate was obtained by Hofer,  Wysocki and Zehnder 
in [HWZ].  It seem that the third part of the bubbling in the contact case, 
especially, the part
concerning the behavior of the connecting tubes along the non-compact
${\bf R}$-direction was not addressed before. We emphasize that in oder
to get the desired compactification without introducing unstable trivial
connecting maps, it is crucial
to know that the "connecting" tube along the non-compact ${\bf R}$-direction
  behaves essentially like the trivial connecting map at $C^0$ sense.
Most analytic 
 part of this
paper  is aimed  to establish the second and the third part of the bubbling
process.

Once  the above bubbling process is established, the main difficulty
to establish the compactification of the moduli space is more conceptual
rather than technical.  In fact what we need here is a 
 a  right definition of stable
maps in the contact case, which should incorporate those symmetry
splitting mentioned above as well as "hidden" symmetries in 
each component of the target (See Sec. 3). In particular, according to
the consideration in Sec 3,
because of these  "hidden" symmetries, one should count the 
${\bf R}$-symmetry of
 each
 component of the target as many times as the number of the 
connected components of the domain lying in the component of the target.  
This will lead to a somewhat "strange" definition of the quivalence
of stable maps in the contact case. 
 
In symplectic geometry,  historically, the
compactness theorem for pseudo-holomorphic maps or connecting $(J, H)$-maps
was first  proved by Gromov and Floer [G, F] by adding certain degenerate maps, 
called
cuspidal maps. Later a smaller compactification was found by using stable maps,
which plays a important role for the recent development in symplectic geometry
(see, for example, [LiT], [FO] and [LT]).
Technically, there is not much difficulty to pass from the cuspidal map
compactification to stable map one. The key is to carefully keep track all
marked points naturally introduced in the bubbling, then to study the
 deformation of the domain equipped with these marked points in a proper moduli
space of curves. In the same vein, the key to get a right compactification
in our case is first to understand the two crucial points mentioned in last
paragraph, then to keep track  carefully 
all marked points and marked lines in the domain,   marked sections in 
the target
naturally appeared in the bubbling  
 and to study the deformation of 
such a structure. Once the desired compactification is established,
our main result about the virtual co-dimension of the boundary
will be a  consequence of
 the  compactness theorem, the index formula proved in [L3]
 and a direct dimension
counting argument.

As mentioned above,  it has been believed that 
bubbling for 
 pseudo-holomorphic maps in symplectization is  a co-dimension
one phenomenon. This has been considered as a major difficulty to
 establish various  
"simple"  and "elementary" constructions, such as   Floer 
homology and G-W invariants,  in contact geometry. A very interesting
 and much more advanced
 construction were
proposed  by Eliashberg, Hofer and Givental 
under
the name contact homology or contact Floer homology (see [E]).

On the other hand, the work of this paper and [L1], [L3]  suggest a rather
different picture on the boundary behavior of the moduli space in contact
geometry. This 
opens the door to construct those "simple" constructions, such
as Floer homology and G-W invariants in contact geometry, which have
essentially same algebraic structures as the ones in symplectic geometry.
It also makes it possible
to generalize various important constructions in symplectic geometry.
We now briefly mention some of these possible applications, which
are outlined in the Sec. 5.

The first application is to define an analogy of  Floer homology in contact
geometry. To
distinguish our construction with the one in [E], which is  under the name
 contact homology or contact Floer homology, 
we call our construction
 index homology. The most natural  way to do this is  to use
 the closed orbits of the Reeb field to generating a chain group
and to count the pseudo-holomorphic maps connecting two closed orbits to
define the boundary map.   As mentioned above, 
 the co-dimension  of the component
of 
broken connecting pseudo-holomorphic maps is   one, and  bubbling is a 
co-dimension two
phenomenon.  Therefore, we are in the exactly the same
 situation   as the usual Floer 
homology, and the desired  index homology can be established as an invariant
of the contact structure. Once this is done, one can also construct 
G-W invariants and use them to define ring structure and the action of the usual
homology of the contact manifold on the index homology, which are the analogies
in  the usual quantum cohomology. To see that the 
index homology so defined is not always trivial,
we introduce Bott-type index homology as a computational tool. Using the Bott-type
homology, we can compute the index homology for a contact manifold, which appears
as a regular zero locus of some local Hamiltonian function which generates a 
$S^1$ Hamiltonian action. It turns out that the index homology of the contact
manifold in this case is just the infinite copies of the usual homology
of its symplectic quotient indexed by the periods of the closed orbits. 
Of course the non-vanishing of index homology implies the Weinstein conjecture.
Therefore, as a corollary, we proved the Weinstein conjecture in
above  case.

%In the above program, we use the moduli space of pseudo-holomorphic
%maps to define the desire index homology theory. On the other hand, in the
%symplectic case, the  usual Floer homology is defined by means of the 
%the perturbed pseudo-holomorphic
%maps with some Hamiltonian perturbations. It turn out that we can define
%another analogy of Floer homology and its pants product in contact case by 
%using perturbed pseudo-holomorphic
%maps with some suitable Hamiltonian perturbations in the symplectization.
% A modified version of this paper and [L1] give  the desired 
%properties on the  moduli space of the  perturbed pseudo-holomorphic
%maps. We will call the homology defined this way the Floer homology in the 
%symplectization.

It is also possible to to use the moduli space differently to define various 
versions of index homology. In particular, in Sec 5, we will outline how
to define an additive quantum homology of a contact manifolds and relative
quantum homology of a symplectic manifold with contact type boundary.

There are also some other important constructions that can be generalized.
%In [V], Viterbo define relative Floer homology for a compact symplectic
%manifold with a convex contact type boundary. Using our work, we can
%generalize this to any compact symplectic manifold with contact type 
%boundary without the assumption about convexity. 
For example,
the relative G-W invariant and its gluing formula can
be established in general, which was  developed by Li-Ruan in [LiR]
before with an extra assumption on the existence of some local $S^1$
Hamiltonian action. 

This paper is organized as follows.

In Sec. 2, we will collect and prove some basic facts about the 
first part of
bubbling.
Almost all of statements there are well-known due to the work
of Hofer and his collaborators.  
However, for the completeness, we give details of the proof for  most
of these
statements. Besides several technical lemmas in section, 
 the most important thing in section is the 
introduction of Hofer's  energy function, which
leads to  the important  notion of finite energe plane in [H].

In Sec. 3, we formulate the notion of stable maps in the symplectization
and the weak-topology of the moduli space of such maps. We then proved
the compactness of the moduli space and the statement concerning the
 co-dimension of its boundary, modulo the statement concerning the exponential
decay of a bubble approaching its non-removable singularity and the statement
concerning the behavior of the "connecting tube". Both of these statements
are proved in Sec. 4.

The last section, Sec 5, is an outline of some possible applications.
the detail of these applications will appear in forth coming papers.

\noindent {\bf Acknowledgment}: The author is very grateful
to Professor G. Tian for  valuable  and inspiring
discussions, for his help on various aspects of the project and
 for his 
encouragement.  

\section{Bubbling}
Let  $(M^{2n+1}, \xi)$ be a contact manifold. This means that $\xi$ is a 
generic $2n$-dimensional subbundle of $TM$. A contact form
$\lambda=\lambda_\xi$ associated to $\xi$ is a 1-form such that
$\lambda\wedge(d\lambda)^n \not = 0$ and $\xi = ker\lambda.$ The 2-form
$d\lambda$ is non-degenerate when restricted to $\xi$ and has a 
1-dimensional kernel at each tangent space of $M$. We denote by $\eta$ the line
bundle generated by $ker(d\lambda)$. It has a canonic section
$X_\lambda$ defined by requiring that $\lambda(X_\lambda) =1$. Since
$\xi\cap\eta=\{0\}$, we have $TM=\xi\oplus\bf{R}X_\lambda.$ Let $\pi:
TM\rightarrow \xi$ be the projection to the first summand.

\vspace{3mm}

\noindent $\bullet$ {\bf Symplectization}:

\vspace{3mm}

The symplectization of $(M^{2n+1}, \xi,\lambda)$ is defined as follows. 

Let ${\widetilde M}$ be $ M\times {\bf R}$ equipped with the exact symplectic
form $\omega=d(e^r \cdot\lambda), $ where $r$ is the coordinate for
the ${\bf R}$-factor. Since $d\lambda$ is symplectic along $\xi$, there
exists a $d\lambda$-compatible almost complex structure $J$ defined on
$\xi$. In fact, the set of all such $J$'s is contractible. We extend $J$
to an $r$-invariant almost complex structure ${\tilde J}$ by
requiring:
$${\tilde J}(\frac{\partial}{\partial{r}})=X_\lambda, \, {\tilde
J}(X_\lambda)=-\frac{\partial}{\partial r}, \mbox{ and } {\tilde
J}=J$$ along $\xi.$

\vspace{3mm}

\noindent $\bullet$
{\bf Equation for ${\tilde J}$-holomorphic curves in ${\widetilde M}$}

\vspace{3mm}

Let ${\tilde u}=(u, a):\Sigma=S^1\times{\bf R}\rightarrow{\widetilde M}$
be a ${\tilde J}$-holomorphic map where $u:\Sigma\rightarrow M$ and
$a:\Sigma\rightarrow{\bf R}$. Then we have 
$$
{\tilde J}({\tilde u})\circ d{\tilde u} = d{\tilde u}\circ
i,\label{star}\eqno(\star)
$$
where $i$ is the standard complex structure on $\Sigma, $ i.e.
$i(\frac{\partial}{\partial s})=\frac{\partial}{\partial t}$ and
$i(\frac{\partial}{\partial t})=-\frac{\partial}{\partial s}.$ Here $(s,
t)$ is the cylindrical coordinate of ${\bf R}\times S^1$. 

Equation ($\star$)\ref{star} is equivalent to the following equations:
$$
\left\{
\begin{array}{cr}
  \pi(u) du+J(u)\pi (u) du\circ i = 0 & \quad \quad\quad(1) \\
 (u^*\lambda)\circ i=da & \quad\quad \quad(2) 
\end{array}
\right.
$$
Equation (1) is equivalent to :
$$\pi(u)(\frac{\partial u}{\partial s}) + J(u)\pi(u)(\frac{\partial
u}{\partial t})=0.\quad \quad\quad \quad(1')
$$

\begin{lemma}\label{lemma1}
$\Delta a =\frac{\partial^2a}{\partial s^2}+\frac{\partial^2 a}{\partial
t^2} \geq 0$ if ${\tilde u} $ is ${\tilde J}$-holomorphic. 
\end{lemma}

\noindent {\bf proof:}
It follows from (2) that 
\begin{eqnarray*}
u^*(d\lambda) &  = & -d(da\circ i)\\
 & = & d(-\frac{\partial a}{\partial t}ds+\frac{\partial a}{\partial
s}dt)\\
 & = & (\frac{\partial ^2 a}{\partial t^2}+\frac{\partial ^2}{\partial
s^2})
ds\wedge dt.
\end{eqnarray*}
Now 
\begin{eqnarray*}
u^*(d\lambda) & = & d\lambda (\pi(\frac{\partial u}{\partial s}), \pi
(\frac{\partial u}{\partial t}))ds\wedge dt\\
& = & d\lambda (\pi (\frac{\partial u}{\partial s}),
J\cdot\pi(\frac{\partial u}{\partial s}))ds\wedge dt\\
& = & g_J(\pi(\frac{\partial u}{\partial s}), \pi (\frac{\partial
u}{\partial s}))ds\wedge dt,
\end{eqnarray*}
where $g_{J}$ is the Riemannian metric 
defined on $\xi$ associated with $d\lambda $ and $J$. Therefore, 
$$\Delta a = |\pi(\frac{\partial u}{\partial s})|^2_{g_{J}} \geq 0.$$

\QED

\vspace{3mm}

\noindent $\bullet$
{\bf Energy}

\vspace{3mm}

Let $\phi\in C^\infty({\bf R}, [\frac{1}{2}, 1])$, $\phi^\prime \geq 0.$ For
any ${\tilde J}$-holomorphic curve ${\tilde u}$, Its $\phi$-energy is defined
as follows:
$$E_\phi({\tilde u})={\int\int}_{{\bf R}^1\times S^1}{\tilde u}^*
d(\phi\lambda)$$ and its energy
$$E({\tilde u}) = \sup_{\phi} E_\phi({\tilde u}).$$

Let $$E_\lambda({\tilde u})={\int\int}_{{\bf R}^1\times S^1}{\tilde u}^*
d(\lambda)$$

Note that:
\begin{eqnarray*}
{\tilde u}^*(d(\phi\lambda)) & = & {\tilde u}^*(d\phi\wedge \lambda +
\phi d\lambda)\\
& = & \phi^\prime(u)da\wedge u^*\lambda+\phi(a)u^*(d\lambda)\\
& = & \{\phi^\prime(a)\{\frac{\partial a}{\partial
s}\lambda(\frac{\partial u}{\partial t})-\frac{\partial a}{\partial
t}\lambda (\frac{\partial u}{\partial
s})\}+\phi(a)d\lambda(\frac{\partial u}{\partial s}, \frac{\partial
u}{\partial t})\}ds\wedge dt\\
& = &\frac{1}{2}\{\phi^\prime(a)\{(\frac{\partial a}{\partial
s})^2+(\frac{\partial a}{\partial t})^2 +  \lambda(\frac{\partial
u}{\partial s})^2+\lambda(\frac{\partial u}{\partial t})^2\}\\
& & +  \phi(a)\{
|\pi(\frac{\partial u}{\partial s})|^2+|\pi(\frac{\partial u}{\partial
t})|^2\}\}ds\wedge dt.
\end{eqnarray*}
This implies that $E({\tilde u})\geq 0.$ 
 
Note: the above local expression ${\tilde u}^*(d(\phi\lambda))$ is valid
for any conformal coordinate. 

\noindent{\bf Example}

Let $x:S^1\rightarrow M$ be a closed orbit of Reeb field $X_\lambda $ of
period $c=\int_{S^1}\lambda(\dot x(t))dt.$ We get a trivial ${\tilde J}$-
holomorphic map ${\tilde u}(s, t)=(u, a) =(X(t), c\cdot s). $ Then 
\begin{eqnarray*}
E_\phi({\tilde u}) & = & \frac{1}{2}\int_{{\bf R}^1\times
S^1}\phi^\prime(a)\{(\frac{\partial a}{\partial
s})^2+\lambda(\frac{\partial u}{\partial t})^2\}ds\wedge dt\\
& = & c^2\int^{\infty}_{-\infty}\phi^\prime(c\cdot s)ds\\
& = & c\{\phi(\infty)-\phi(-\infty)\}\\
& =& \frac{1}{2}c.
\end{eqnarray*}

\vspace{3mm}

\noindent $\bullet$
{\bf Bubbling}:

\vspace{3mm}

\begin{lemma}\label{bubbling}

Let $(X, d)$ be a complete metric space and $\phi:X\rightarrow {\bf
R}^{+}=[0,\infty)$ be a continuous function. Given $x\in X$ and $\epsilon>0$,
there exists $x^\prime\in X $and $\epsilon^\prime >0$ such that 

(1) $\epsilon^\prime \leq \epsilon, \, \phi(x^\prime)\epsilon^\prime
\geq \phi(x)\cdot \epsilon;$

(2) $d(x, x')\leq 2\epsilon$;

(3) $ 2\phi(x')\geq \phi(y)$ for all $y\in X$ such that $d(y, x')\leq
\epsilon^\prime.$

\end{lemma}

The proof is elementary. See [H-V].

\begin{pro}

Let ${\tilde u}_n =(u_n, a_n):{\bf R}^1\times S^1$ (or ${\bf
C}$)$\rightarrow {\widetilde M}$ be a sequence of ${\tilde J}$-holomorphic
maps such that (i) there exists a constant $c>0$ such that $E({\tilde
u}_n)<c$; (ii) for each ${\tilde u}_n$, there exists a $x_n\in {\bf
R}^1\times S^1$ such that $|du_n(x_n)|\rightarrow \infty.$ Then the
sequence $\{{\tilde u}_n\}^\infty_{n=1}$ will bubble off at
$\{x_n\}^\infty_{n=1}$ a bubble ${\tilde v}:{\bf C}\rightarrow {\widetilde
M}$, which is ${\tilde J}$-holomorphic such that

(a) $|d{\tilde v}(0)|=1$;

(b) $|d{\tilde v}(y)|\leq 2$ for any $y\in {\bf C};$ and

(c) $E({\tilde v})<c. $
\end{pro}

\noindent{\bf Proof}:

Apply Lemma \ref{bubbling} to the case that $(X, d)={\bf R}^1\times
S^1$, $\phi=|d{\tilde u}_n|$, $x=x_n$, and $\epsilon=\epsilon_n$, where
$\{\epsilon_n\}^\infty_{n=1}$ is a sequence such that $d_n\cdot
\epsilon_n = |d{\tilde u}_n(x_n)|\cdot\epsilon_n\rightarrow\infty. $ We
may assume that 

(a) $|d{\tilde u}_n(x_n)|\cdot\epsilon_n\rightarrow \infty;$

(b) $2|d{\tilde u}_n(x_n)|>|d{\tilde u}_n(y)|$ for any $y\in
D_{\epsilon_n}(x_n);$

(c) ${\tilde u}_n(x_n)\in M\times \{0\}$ after a translation in ${\widetilde
M}$. 

Fix $R>0$, define ${\tilde v}_{n, R}:D_{R}\rightarrow{\widetilde M}$ 
to be ${\tilde v}_{n, R}(x)={\tilde u}_n(x_n+\frac{x}{d_n}).$

Note that when $n$ is large enough, $d_n\cdot\epsilon_n>R.$ Hence,
$x_n+\frac{x}{d_n}\in D_{\epsilon_n}(x_n)$ for $x\in D_{R}$. This
implies that 

(i) $|d{\tilde v}_{n, R}(0)|=\frac{1}{d_n}|d{\tilde
u}_n(x_n)| =1;$

(ii) $|d{\tilde v}_{n, R}(x)| =\frac{1}{d_n}|d{\tilde
u}_n(x_n+\frac{x}{d_n})|\leq \frac{2d_n}{d_n}=2$ for any $x\in D_R;$

(iii) ${\tilde v}_{n, R}(0)\in M\times \{0\}.$ 

Now the standard elliptic estimation implies that ${\tilde v}_{n, R}$
is $C^\infty$-convergent to a ${\tilde J}$-holomorphic map ${\tilde v}_R:$ 
$D_R\rightarrow {\widetilde M}$ after taking a subsequence of
${\tilde v}_{n, R}$. Let $R_n\rightarrow \infty$, and taking a diagonal
subsequence, ${\tilde v}_{n, R_n}$ is $C^\infty$-convergent to a
${\tilde J}$-holomorphic map ${\tilde v}=\cup_{R}{\tilde v}_R:{\bf
C}\rightarrow{\widetilde M}$ such that (a) $|d{\tilde v}(0)|=1$; (b)
$|d{\tilde v}(x)|\leq 2, x\in {\bf C};$ (c) $E({\tilde v})<c;$
(d) ${\tilde v}(0)\in M\times\{0\}.$

\QED

The following lemma will be used to prove that the bubbling will stop
after finite steps. 

\begin{lemma}\label{lemma3}

Fix $c>0$. Let $V$ be the collection of all ${\tilde J}$-holomorphic
maps ${\tilde v}:{\bf C}\rightarrow{\widetilde M}$ satisfying the properties
(a)-(c) in the previous proposition. Then there exists a constant
$\epsilon >0$ such that for any ${\tilde v}\in V$, $\int_{\bf
C}{\tilde v}^*(d\lambda)>\epsilon.$
\end{lemma}

\noindent{\bf Proof}:

If not, there would exist ${\tilde v}_n:{\bf C}\rightarrow {\widetilde M}$ of
${\tilde J}$-holomorphic maps such that (a) $|d{\tilde v}_n(0)| =1; $
(b) $|d{\tilde v}(x)|\leq 2, x\in {\bf C};$ (c) $E({\tilde v}_n)<c;$ and
(d) ${\tilde v}_n(0)\in M\times\{0\}$ after ${\bf R}$-translation in
${\widetilde M}$; (e) $\lim_{n\rightarrow\infty}\int_{\bf C}{\tilde
v}^*_n(d\lambda) =0$. 

Now (a)-(d) implies that ${\tilde v}$ is locally $C^\infty$-convergent
to a ${\tilde J}$-holomorphic map ${\tilde v}:{\bf C}\rightarrow{\widetilde
M}$ with same properties of (a)-(d). Now 
$$\int_{\bf C}{\tilde
v}^*(d\lambda)=\lim_{R\rightarrow\infty}\int_{D_R}{\tilde
v}^*(d\lambda)=\lim_{R\rightarrow\infty}\lim_{n\rightarrow\infty}\int_{D_R}{\tilde
v}^*_n(d\lambda)=0.$$
It follows from Lemma \ref{4prime} that ${\tilde v}$ is a constant map.
This contradicts to (a). 

\QED

\begin{lemma}\label{lemma4}

Let ${\tilde u}:{\bf R}^1\times S^1\rightarrow{\tilde M}$ be a ${\tilde
J}$-holomorphic map such that $E({\tilde u})<\infty$ and $\int_{{\bf
R}^1\times S^1}{\tilde u}^*(d\lambda)=0.$ Then either ${\tilde u}$ comes
from a closed orbit of $X_\lambda$ as in Example 1, or ${\tilde u}$ is a
constant map. 
\end{lemma}

\begin{lemma}\label{4prime}

Let ${\tilde u}:{\bf C}\rightarrow{\tilde M}$ be a ${\tilde J}$-
holomorphic map such that $E(\tilde u)<\infty$, and $\int_{\bf C}{\tilde
u}^*(d\lambda)=0$, then ${\tilde u}$ is a constant map. 
\end{lemma}

\noindent{\bf Proof} of Lemma \ref{lemma4}

\begin{eqnarray*}
0 &= & \int_{{\bf R}^1\times S^1}{\tilde u}^*(d\lambda) =
\frac{1}{2}\int_{{\bf R}^1\times S^1}\{|\pi(\frac{\partial u}{\partial
s})|^2 +|\pi(\frac{\partial u}{\partial t})|^2\}ds\wedge dt\\
& \Longrightarrow & \pi(\frac{\partial u}{\partial
s})=\pi(\frac{\partial u}{\partial t})=0\\
& \Longrightarrow & u\mbox{ is tangent to } {\bf R}X_\lambda.
\end{eqnarray*}
This implies that 
$ u=x\circ f,$ where $ f:{\bf R}^1\times {\bf R}^1\rightarrow {\bf R}$ and
$x=x(t)$ is the solution of $\frac{dx}{dt}=X_\lambda(x(t)).$ Here we treat
$u$ as a function defined on ${\bf R}^1\times {\bf R}^1$ which is periodic
in the second variable.

Therefore, 
$$
\left\{
\begin{array}{l}
\frac{\partial u}{\partial s}  =  \dot x\frac{\partial f}{\partial s}
= \frac{\partial f}{\partial s}X_\lambda(f)\\
\frac{\partial u}{\partial t}  =  \dot x\frac{\partial f}{\partial t}
 = \frac{\partial f}{\partial t}X_\lambda(f).
\end{array}
\right.
$$

This implies that $\lambda(\frac{\partial u}{\partial s})=f_s$ and
$\lambda(\frac{\partial u}{\partial t})=f_t$. Now the equation
$\lambda\circ du=-da\circ i$ is equivalent to 
$$
\left\{
\begin{array}{l}
\lambda(\frac{\partial u}{\partial s})=-a_t\\
\lambda(\frac{\partial u}{\partial t})=a_s.
\end{array}
\right.
$$
We have 
$$
\left\{
\begin{array}{l}
a_s=f_t\\
a_t=-f_s.
\end{array}
\right.
$$
That is $F=a+fi$ is holomorphic on ${\bf  C}$.  Note that here 
we treat
$a$ as a function on ${\bf  R} \times {\bf  R}$ which is periodic on
the second variable.
Therefor, $F'=\frac{\partial F}{\partial z}$ is also holomorphic. 

Now $|dF|^2=|\frac{\partial F}{\partial z}|^2 =|\frac{\partial
a}{\partial z}|^2+|\frac{\partial f}{\partial z}|^2.$ If $|da|$ is
bounded, then $|df|$ is also bounded.  The holomorphic funtion
$F'$ defined on ${\bf C}$ has bounded norm, hence, is a constant. 
This implies that 
$$F=c\cdot z +d
=(c_1+c_2i)\cdot(s+ti)+d_1+d_2i=(c_1s-c_2t+d_1)+(c_1t+c_2s+d_2)i.$$
Hence $f= (c_1t+c_2s+d_2)$ and  $a(s,t)= c_1s-c_2t+d_1$.
Sinece $a$ is  periodic in $t$: $a(s, t+1)=a(s, t).$ This implies that
$a=c_1\cdot s+d_1.$ 

We claim that in this case 
$x(t)$ is  a closed orbit of $X_\lambda$,
Suppose this is not true, 
$f$ has to be periodic in $t$: $f(s, t+1)=f(s, t).$ This implies that
$f=c_2\cdot s+d_1.$ Now 

But $$ 
\left\{
\begin{array}{l}
a_s=f_t=0\\
a_t=-f_s=-c_2
\end{array}
\right.
$$
 implies that $c_1=c_2=0,$ and hence $F$ is constant. 

Therefore, we may assume that $|da|$, hence $|d{\tilde u}|$ is not
bounded. Then the bubbling process described before is applicable to
this case and will produce a bubble ${\tilde v}:{\bf
C}\rightarrow{\widetilde M}$ with the properties that (a) $E({\tilde
v})<\infty;$ (b) ${\tilde v}^*(d\lambda)=0;$ (c) $|d{\tilde v}(0)|=1$;
(d) $|d{\tilde v}|$ is bounded. But we will prove in Lemma 5 that (a) and (b)
imply that ${\tilde v}$ is a constant map. This contradicts with (c). 

\QED

\noindent{\bf Proof of Lemma \ref{4prime}}

As in Lemma \ref{lemma4}, we have $u=x\circ f$ and $F=a+fi$ is
holomorphic. If $|df|$ and hence $|d{\tilde u}|$ is unbounded, as above,
we would have a bubble ${\tilde v}=(v, a)$ with the properties (a)-(d)
above. Then (b) implies that $v=x_1\circ f_1$ and $F_1=a_1+f_1i$ are
holomorphic. Now (d) implies that $|da_1|$ and hence $|df_1|$ is
bounded. Therefore $F'_1=\frac{\partial }{\partial z}F_1$ has bounded
norm. Hence 
$$F_1=(c_1s-c_2t+d_1)+(c_1t+c_2s+d_2)i$$ and 
$$
\left\{
\begin{array}{l}
a_1=c_1s-c_2t+d_1\\
f_1=c_1t+c_2s+d_2.
\end{array}
\right.
$$
Now 
\begin{eqnarray*}
{\tilde v}^*(d(\phi\cdot\lambda)) & = 
 & \frac{1}{2}\phi'(c_1s-c_2t+d_1)\{(\frac{\partial a_1}{\partial
s})^2+(\frac{\partial a_1}{\partial t})^2+(\frac{\partial f_1}{\partial
s})^2+(\frac{\partial f_1}{\partial t})^2\}\\
& = & \phi'(c_1s-c_2t+d_1)\{c_1^2+c_2^2\}.
\end{eqnarray*}
If $c_1$ or $c_2\not = 0$ (say $c_1>0$), then

\begin{eqnarray*}
E_\phi({\tilde v}^*) & = & (c_1^2+c_2^2)\int^{\infty}_{-
\infty}\int^\infty_{-\infty}\phi'(c_1s-c_2t+d_1)dsdt\\
& =& \frac{c_1^2+c_2^2}{c_1}\int^\infty_{-\infty}\{\phi(+\infty)-\phi(-
\infty)\}dt = +\infty
\end{eqnarray*}
Hence $c_1=c_2\equiv 0$ and ${\tilde v}=$constant. But this contradicts
with (c).

Therefore, $|df|$ and hence $|da|$ is bounded. Hence $F'$ has constant
norm. We get $F=cz+d$ again. As above $E({\tilde u})<\infty$ implies
that $c=0.$

\QED

\begin{pro}\label{pro2}

Let ${\tilde u}:{\bf R}^1\times S^1$ (or ${\bf C}$) $\rightarrow
{\widetilde M}$ be a ${\tilde J}$-holomorphic map such that $E({\tilde
u})<\infty.$ Then there exists a $c>0$ such that $|d{\tilde u}(x)|<c.$
\end{pro}

\noindent{\bf Proof}:

$E({\tilde u})<c\Longrightarrow \int_{{\bf R}^1\times S^1}{\tilde
u}^*(d\lambda)< c'> 0.$

If $|d{\tilde u}|$ is not uniformly bounded, then there exists
a sequence $x_n=(s_n,
t_n)$ with $s_n\rightarrow \pm\infty$ such that $|d{\tilde
u}(x_n)|\rightarrow \infty.$ This will produce a bubble ${\tilde v}$
with the properties (a)-(d) as in Lemma \ref{lemma4}. Note that (b)
${\tilde v}^*(d\lambda)=0$ follows from the fact that the $s$-coordinate
of $x_n=(s_n, t_n)$ tends to $\pm\infty$. As in Lemma \ref{lemma4}, this
leads to a contradiction. The proof for the case that ${\tilde u}:{\bf
C}\rightarrow{\tilde M}$ is the same. 

\QED

Now assume that the contact $1$-form $\lambda$ is generic so that $1$ is
not an eigen value of the Poincare \u are returning map at any closed orbit
of $X_\lambda.$ This implies that the set of unparameterized closed orbits of
$X_\lambda$ are discrete. 

\begin{pro}
Let ${\tilde u}:{\bf R}^1\times S^1\rightarrow {\widetilde M}$ be a ${\tilde
J}$-holomorphic map with $E({\tilde u})<\infty$ and ${\tilde u}\not =$
constant map. Then $\lim_{s\rightarrow \infty}{\tilde u}(s, t)$, when
being projected to $M$ is either a closed orbit of $X_\lambda$ or a
constant map. Assuming the first case happens, then ${\tilde u}(s, t)$
is convergent to two closed orbits $x_{\pm}$ asymptotically with a
exponential decay rate. 
\end{pro}

\noindent{\bf Proof}:
The proof for the part concerning the exponential decay of the last 
statement is given in Sec. 4.

By proposition \ref{pro2}, there exists a $C>0$ such that $|d{\tilde
u}|<C.$ For any fixed $L>0$, we define ${\tilde v}_{n, L}={\tilde
u}(s+n, t):[-L, L]\times S^1\rightarrow{\widetilde M}$. Then ${\tilde v}_{n,
L}$ is $C^\infty$-convergent to ${\tilde v}_L$ after taking a subsequence
and ${\tilde v}={\tilde u}(s+n, t)$ is locally $C^\infty$-convergent to
${\tilde v}_\infty:{\bf R}^1\times S^1\rightarrow{\widetilde M}$ such that
$E({\tilde v}_\infty)<\infty$ and $\int_{{\bf R}^1\times S^1}{\tilde
v}^*(d\lambda)=0.$ Hence ${\tilde v}_\infty =$ constant map or ${\tilde
v}_\infty(s, t)=(x(c\cdot t+d_1), c \cdot s+d_{2}) $ with $\frac{dx}{dt}=X_\lambda(x)$ and
$c=\int_{S^1}x^*\lambda dt.$ Note that in the later case, $a({\tilde v}_n(s,
t))\rightarrow\pm\infty$ as $n\rightarrow \infty.$ We may assume that $c>0$, and
hence $a(s+n, t)\rightarrow +\infty$ as $ n\rightarrow  \infty$.

Assume the second case happens. Applying the same argument to the
negative end of ${\bf R}^1\times S^1$, we get
$\lim_{n\rightarrow\infty}p\circ {\tilde u}(s-n, t)|_{[-L, L]\times S^1}
= x_{-}(c_{-}t+d_{-})$ for some closed orbits $x_{-}$ of $X_\lambda$ of
period $c_{-},$  or  a constant map. Here $p$ is the projection ${\widetilde M}\rightarrow M.$
Assume again that it is not the constant map. 

Now
$$\int_{{\bf R}^1\times S^1}{\tilde
u}^*(d\lambda)=\lim_{n\rightarrow\infty}\int_{\{n\}\times S^1}{\tilde
u}^*(\lambda)-\int_{\{-n\}\times S^1}{\tilde u}^*(\lambda)=c_+ -c_-.$$

If ${\tilde u}(s+n_i, t)|_{[-L, L]\times S^1}, n_i\rightarrow\infty$ is
any other convergent sequence, then the limit must also be a closed
orbit of period $c_+$. 
Let $x'$ be the closed orbit. Under the assumption
that $\lambda$ is generic, there are only finite $c$-period closed orbits
of $X_\lambda$. If $x\not =x'$, we can find $(s_i, t_i)\in {\bf
R}^1\times S^1$ with $s_i\rightarrow +\infty$ and ${\tilde u}(s_i,
t_i)\not\in $ a small neighborhood of the set of $c$-closed orbits. Then
${\tilde u}(s+s_i, t)|_{(-L, L)\times S^1}$ is $C^\infty$-convergent to
a constant map. This implies that $\int_{{\bf R}^1\times S^1}{\tilde
u}^*(d\lambda) = -c_-$ and leads to a contradiction. 

Therefore, we conclude that 
$x$ and $x'$ are the same as unparameterized curves.
Then it t is easy to see that  as parametrized curve there is also only one
limit
$\lim_{s\rightarrow\pm\infty}p\circ{\tilde
u}(s, t) =x_{\pm}(c_\pm t +d_\pm).$ 

 In the case of the above limits are closed orbits,
$\lim_{s\rightarrow\pm\infty}a(s, t)=\pm\infty$. This can be seen easily
from the explicit expression of the limit of  local convergence of 
the sequence ${\tilde v}_{n, L}$ introduced at the beginning of the proof.

\QED

\section{Compactness}

Let $M_\pm$ be the two ends of ${\tilde M}=M\times {\bf R}$. We 
consider subset of all finite energy ${\tilde J}$-holomorphic maps whose
two ends asymptotically approximate to two closed orbits in $M_\pm$.
More precisely, given  two  parametrized closed orbits 
$x_\pm:S^1\rightarrow
M_\pm\simeq M$, let $\{x\}_\pm$ be the set of all such parametrized closed 
orbits differ from $x_\pm $ by $S^1$ actions. Define 
$${\widetilde{\cal M}}(x_-, x_+, {\tilde J})=\{{\tilde u}\, |\, 
{\tilde u}:{\bf R}^1\times S^1\rightarrow{\tilde M}, 
{\bar\partial}_{\tilde J}=0, \lim_{s\rightarrow\pm\infty}u(s, t)=x'_\pm(t)
, x'_\pm\in \{x_\pm\}\}.$$
There is an obvious 3-dimensional symmetry 
group acting on the moduli space. The actions are induced from the ${\bf R}$-
translations on the target ${\tilde M}$ and ${\bf R}^1\times S^1$-action
on the domain ${\bf R}^1\times S^1$.   Note that the effect of 
the two types  of actions induced from  ${\bf R}$-actions on 
the target and  the domain are
 never identical
 unless they act on the trivial
${\tilde u}={\tilde u}(s, t)=(x_\pm(t), s).$

Let ${\cal M}(x_-, x_+, {\tilde J})={\widetilde{\cal M}}(x_-, x_+;
{\tilde J})/{{\bf R}^2\times S^1}.$ 

\vspace{3mm}

\noindent $\bullet$
{\bf Energy}:

\vspace{3mm}

Given ${\tilde u}\in {\widetilde {\cal M}}(x_-, x_+; {\tilde J}), $
$$\int_{{\bf R}^1\times S^1}{\tilde u}^*(d\lambda)=\int_{\partial({\bf
R}^1\times S^1)}{\tilde u}^*(\lambda)=\int_{S^1}x^*_+(\lambda)-
\int_{S^1}x^*_-(\lambda)=c_+-c_-,$$ 
where $c_\pm$ are the periods of $x_\pm$. 

\begin{lemma}

Given $u\in{\widetilde{\cal M}}({\tilde x}_-, {\tilde x}_+; {\tilde
J})$, then $\int_{{\bf R}^1\times S^1}{\tilde u}^*(d\lambda)\geq 0$ and
equality holds if and only if ${\tilde x}_-={\tilde x}_+$ and ${\tilde
u}(s, t)=x_\pm(t).$ 
\end{lemma}

We will call such ${\tilde u}$ trivial map. Therefore, if $x_-\not=x_+$,
${\cal M}(x_-, x_+; {\tilde J})$ does not contain trivial map and the
${\bf R}^2$-action is free. 

\vspace{3mm}

Given ${\tilde u}\in {\widetilde{\cal M}}(x_-, x_+; {\tilde J})$,
\begin{eqnarray*}
E_\phi(\tilde u) &= &\int_{{\bf R}^1\times S^1}{\tilde
u}^*(d(\phi\lambda))=\int_{\partial({\bf R}^1\times S^1)}{\tilde
u}^*(\phi\lambda)\\
& = & \int_{S^1}\phi(a(+\infty))x_+^*(\lambda)-\int_{S^1}\phi(a(-
\infty))x_-^*(\lambda)\\
& = & c_+-\frac{1}{2}c_-\geq c_+-\frac{1}{2}c_+=\frac{1}{2}c_+>0
\end{eqnarray*}

\vspace{3mm}

{\bf Compactification of ${{\cal M}}(x_-, x_+; {\tilde J})$}:

$\bullet$  Stable ${\tilde J}$-map connecting $x_-$ and $x_+$:

There are two different ways to define this notion. One is 
saver but gives less information.  We start with this saver one
first. The Remark 3.2 in this section will tell us how to modify
the definition here to get the  more informative one.

Domain $\Sigma =\Sigma_{\tilde u}$ of a stable ${\tilde J}$-map 
${\tilde u}$ connecting  $x_-$ and $x_+$ can be written as
$\Sigma=\cup_i\Sigma_{{ p}_i}\cup_j\Sigma_{{ b}_j}$, $i=1,
\cdots, P$, and $j=1, \cdots, B, $ of the union of domains $\Sigma{
p}$ of its principal components and $\Sigma_{ b}$ of its bubble
components. Each $\Sigma_{ p}$ or $\Sigma_{ b}$ is
holomorphically equivalent to $S^2$. As a curve, $\Sigma$ is semi-
stable. This means that the worst singularity of $\Sigma$ is double
point singularity. The components of $\Sigma$ form a connected tree. There 
are two particular marked points $-\infty$ on $\Sigma_{{ p}_1}$ and
$+\infty$ on $\Sigma_{{ p}_P}.$ on each $\Sigma_{{p}_i}$, there
are double points $d_{i, -}$ and $d_{i, +}$ such that $\Sigma_{{
p}_i}$ and $\Sigma_{{ p}_{i+1}}$ are jointed together in $\Sigma$ at
the double point $d=d_{i, +}=d_{i+1, -}$. Therefore, the domain of
principal component forms a chain. These joint double points 
$d_i=d_{i,+}=d_{i+1, -}$ are  divided into two classes according
 to the asymptotic behavior of $u$ when $u$ approaches  $d_i$. 
We will use $I_P$ to denote the set of those indices $i $ such that $u$
approximates to some closed orbit $x_{p_i}$ when it approaches $d_i,$
  while for the other  $i\in P\setminus I_{P}$, ${\tilde u}$ is well defined at
 $d_{i, +}= d_{i+1, -}.$ Similarly, for all other double points of 
$\Sigma,$ we will make such a distinction. For each of the double point
which is treated as infinity of an end, we will introduce a fix
$S^1$-parameterization at the infinity of the end. We will include this
as part of the structure of $\Sigma$. This can be done, for example,
by identify a small neighborhood  $U$ of some double point $d$ with
two copies of $R^+\times S^1$ and using  the $S^1$-parameterization
on each of $R^+\times S^1$ to give the desired $S^1$-parameterization.
For the later application, we mention the following "canonical"
way to give the $S^1$-parameterization for the double point on each of 
top bubbles in the bubble tree. For each of such bubble, we first add 
a marked point $y$, then choose another marked point z along the circle of 
of the radius 1 centered at $y$. The ray connecting $y$ and $z$ and
started at $y$ gives the required parameterization at infinity.
We remark that it is only the parameterization itself is included in the
 structure of $\Sigma$, not the other things used to define it.
Therefore the dimension of the symmetry group of a top bubble is three.

Note that our definition of the domain of a stble map
is similar to the one used in the usual Gromov-Floer
theory. However if we restricted to the compactification of
${{\cal M}}(x_-, x_+; {\tilde J})$, then the domains of its
elements subject to further restrictions. Although it  does
not effect  constructions in this paper and the subsequent
forth coming papers in any essential way, these furth restrictions simplify the 
possible intersection pattern of domains and make the situation
here is different from the corresponding case in the Gromov-Floer
theory. We refer the readers to Remark 3.2 of this section on this.

The target $U$ of ${\tilde u}$ is a union 
${U}=\cup _{i\in I_P}{\tilde M}_{p_i}$ with each ${\tilde
M}_{p_i}\simeq{\tilde M}$. Note that here we have somewhat abused the 
notation as it may happen that on each ${\tilde M}_{p_i}$, there may
exist more than one ${\tilde u}_{p_i}$'s. Each ${\tilde M}$ has two ends ${\tilde
M}_{p_i, \pm}$ and we identify ${\tilde M}_{p_i, +}$ with ${\tilde
M}_{P_{i+1}, -}$. On each $M_{p_i, \pm}$, there is a particular closed
orbit $x_{p_i,\pm}$ associated to each index $p_i$, $i\in I_P$ and
possibly some other closed orbits $x_{p_i, b_{j,l}}$, $i=1, \cdots, P$.
Here $b_{j,l}$ are indices of the double points on bubble component
$\Sigma_{p_i, b_j}$. Here we have relabeled bubble component $\Sigma_{b_k}$ 
before as $\Sigma_{p_i, b_j}$, where $p_i$ is principal component on
which the bubble $\Sigma_{b_k}$ lies.

Note that $x_-$ lies on the negative end of the first ${\tilde M}_{p_i}$'s 
and $x_+ $ lies on the positive end of the last ${\tilde M}_{p_i}$'s,
$p_i\in I_P.$

The stable map ${\tilde u}=
\cup_{i=1}^P{\tilde u}_{p_i}\cup^B_{j=1}{\tilde u}_{p_i,
b_j}$ such that

(i) ${\tilde u}_{p_i}:\Sigma_{p_i}-\{\mbox{double points}\}\rightarrow{\tilde
M}_{\phi(p_i)}$ and ${\tilde u}_{b_j}:\Sigma_{p_i, b_j}\setminus \{\mbox{double
points}\}\rightarrow{\tilde M}_{\phi(p_i)}$ are ${\tilde J}$-
holomorphic. Here $\phi(p_i)$ is a function from the set of indices
$\{1, \cdots, P\}$ to $I_P$, which is the identity map when being
restricted to $I_P$. 

(ii) Along each end near the double point $d_i\in \Sigma_{p_i}$,
${\tilde u}_{p_i}$ is convergent exponentially to some 
parametrized  periodic orbit $x_{p_i}$,
if $i\in I_P$. Otherwise, ${\tilde u}_{p_i}$ is well-defined at $d_i$ and
 ${\tilde u}$ has
an ordinary double point at $d_i$. Similarly at each double point on
bubble components or double point on principal components other than
these $d_i$'s, ${\tilde u}$ either asymptotically approximates to a closed orbit
$x$ or  it extends smoothly across these double points.  Note that in the case that
${\tilde u}$ asymptotically approximates to a parameterization
 closed orbit along some double point, the $S^1$-parameterization 
(covering) of the closed orbit is given by the $S^1$-parameterization 
of the end.

(iii) On each ${\tilde M}_{p_i}$, $i\in I_P$, there is an ${\bf R}^1$-
action of $r$-translation. We require that the isotropy subgroup of
the components of ${\tilde u}$ in ${\tilde M}_{P_i}$ is not the entire ${\bf
R}^1$. This implies that each ${\tilde M}_{p_i}$ contains at least one
bubble components if the principle component  of ${\tilde u}$ in ${\tilde M}_{p_i}$ 
is a trivial
component. Here a trivial principal component of ${\tilde u}$ in ${\tilde
M}_{p_i}$ is the ${\tilde J}$-holomorphic map $ {\tilde u}_{p_i}:{\bf R}^1\times
S^1\rightarrow{\tilde M}_{p_i}$ such that ${\tilde u}_{p_i}(s, t)=(x(ct), cs+d)$ for some
periodic orbit $x$ in $M$. Clearly the isotropy group of such an ${\tilde u}_{p_i}$ is
${\bf R}^1$ itself. 

(iv) Each constant bubble component is stable in the sense that it
contains at least three double points.

(v) ${\tilde u}$ connects $x_{-}$ and $x_{+}$,  meaning that it connects
some $x'_{-}\in \{x_{-}\}$ and $x'_{+}\in \{x_{+}\}$.

Note that similar to the stable
maps used in the usual Floer homology, there are two kinds of trivial
components, and the trivial principal components play similar role as
closed orbits of a Hamiltonian system regarded as trivial principal
components of a stable $(J, H)$-map in Floer homology.
The reason to
rule out this kind of components can be seen as follows. 

{\bf Example}

Let ${\tilde u}:S^1\times{\bf R}^1\rightarrow{\tilde M}$ be a ${\tilde J}$-
holomorphic map connecting two closed orbits $x_-$ and $x_+$. Assume
that ${\tilde u}$ is not trivial. Hence the effect of the  ${\bf R}$-actions 
on ${\tilde u}$ induced by the  ${\bf R}$-actions on
${\tilde M}$  is different from those induced by  the ${\bf R}$-actions on the domain
$S^1\times {\bf R}^1$. Define ${\tilde u}_n(s, t)={\tilde u}(s+n, t)$.
Then $\{{\tilde u}_n\}^\infty_{n=0}$ is locally $C^\infty$-convergent to
${\tilde u}_\infty={\tilde u}_{\infty, 0}\cup {\tilde u}_{\infty, 1}$
where ${\tilde u}_{\infty, 0}={\tilde u}$ and ${\tilde u}_{\infty,
1}:{\bf R}^1\times S^1\rightarrow{\tilde M}_1$ with ${\tilde u}_{\infty,
1}(s, t)=(x_+(ct), cs+d).$ Iterating this process, we can produce any number of
trivial principal components as limit. 

Note that this example also indicates that even though the energy
$E({\tilde u})$ is a constant for any 
${\tilde u}\in{\widetilde{\cal
M}}(x_-, x_+; J)$, it is not preserved when passing to the limit.
We will see that for the element ${\tilde u}$ in the moduli space
  ${\tilde u}\in{\widetilde{\cal
M}}(x_-, x_+; J)$ of stable connecting
maps, the energy is uniformly bounded. However, without the assumption
of stability, there is no such a bound as above example shows.  On the
other hand, the quantity $\int_{{\bf R}^1\times S^1}{\tilde
u}^*d\lambda$ is obviously preserved under the limit process. 

$\bullet$ {\bf Compactness}

Let $${\overline{\cal M}}(x_-, x_+; J)=\{[{\tilde u}]\,|\, {\tilde
u}\mbox{ is a stable ${\tilde J}$-map connecting $x_-$ and $x_+$; } E({\tilde u})
\mbox{ is finite.}\}.$$ 
Here $[{\tilde u}]$ is the equivalent class of ${\tilde u}$. 

The definition here needs some explanation. In the usual quantum homology
and Floer homology in sympletic geometry, to form the moduli space
${\cal M}(x_-, x_+; J)$ and its compactification, one needs to fix a 
relative homotopy class, which is represented by the elements in these
moduli spaces.  Therefore, there are options here. One is to follow the
usual definition, which is saver but less informative. We will leave
 the rutin 
formulation of this saver definition  to our reader. On the other hand,
the Remark 3.2 together with the next two lemmas imply that. One still
can
prove the compactness without restricting to a particular relative
homotopy class.

\begin{definition}

Two stable
${\tilde J}$-map $\tilde u_{1}$ and $\tilde u_{2}$ connecting $x_-, x_+$ are
 said to be equivalent if
there exists an equivalence $\phi:\Sigma_1\rightarrow\Sigma_2$ of their
domains and an equivalence $\psi: {\tilde U}_1\rightarrow {\tilde U}_2$ of
the liftings of their targets such that $u_1=\psi^{-1}\circ u_2\circ\phi$,
 where $\phi$
is a homomorphism of $\Sigma_1$ and $\Sigma_2$ such that it is bi-holomorphic 
along each of their components and preserves the variable
$t\in S^1$ along the chain of principal components  
and preserves the $S^1$-parameterizations at infinity along those ends
of bubble components or principal components approaching to some closed
orbits,
and $\psi$ is 
induced from ${\bf R}$-translations  on each component of the target $U$, in
 the sense explained in the following.
Here ${\tilde U}_1 $ and ${\tilde U}_2$ are certain finite liftings determined
by 
 the connected components of the bubble tree in each of the components 
of $U$'s.

\end{definition}

Note that the components
of the domain $\Sigma$  of a stable map ${\tilde u}$ forms a  connected
tree. If we fix a component ${\tilde M}_{i}={\tilde M}$ of the target $U$, 
and collect those components
of the domain in the bubble tree above whose images stay in ${\tilde M}_{i}$,
 the components may not be connected anymore. We will
associated to each 
of a connected components $\Sigma_{i, j,} j=1, \cdots, J_{i}$,
 of the domain in ${\tilde M}_{i}$,  a  ${\tilde M}_{i,j}={\tilde M}$.
 We collect all of these ${\tilde M}_{i,j}$ together with same ends as before
as the lifting of $U$ mentioned above. Then the $\psi$ defined above is just
induced by the ${\bf R}$-translations on each component of ${\tilde U}$'s.
In particular, it follows from this definition that on each component of
the target of a stable map, there are as many dimensions of ${\bf R}^1$-actions
as the number of the connected components of the domain in the component
of the target. 

There is a special case that the above definition on connected component of the domain
in a component of  the target is not applicable. According to above
definition, if a component of the target contains a trivial connecting
map, the domain of this map clearly is an isolated component in the
domain of the orignal map inside the component of the target. For the
obvious reason that one can not  assign an extra ${\bf R}^1$-symmetry of the
target associated the the trivial map. However, as we will prove  in
this section that on the component of the target, there exists at least
one non-trivial connected component of the domain. We simply define
a connected component in this case as the union of the non-trivial one
with the domain of the trivial map. In the case that there are several
non-trivial connected components together with several trivial maps in 
one component of the target, we  consider all possible combinations
of them and consider this as part of the data in the definition of
stable map. Another way to deal with this particular case is to only
count the ${\bf R}^1$-symmetry of the domain for each of such stable
component.

With this interpretation, we note that with respect to the symmetry
group so defined, the isotropy group of a stable map is always finite, 
which is important for defining the virtual moduli cycles in [L1] and
will be proved there.

\begin{theorem}
${\overline{\cal M}}(x_-, x_+; J)$ is compact and Hausdorff with
respect to the $C^\infty$-weak topology, which is a compactification of
${\cal M}(x_-, x_+; J)$. 
\end{theorem}

{\bf Proof}:

The Hausdorffness follows from the stability of elements in ${\bar{\cal
M}}(x_-, x_+; J)$. The proof of the corresponding theorem in [LT] can be
easily adapted here. We refer our readers to the proof there. 

\begin{lemma}\label{lemma5}

There exists a constant $N=N(x_-, x_+)$ such that for any ${\tilde
u}\in[{\tilde u}]\in{\overline{\cal M}}(x_-, x_+; J)$, the number of
components of $u$ is less than $N$. 
\end{lemma}

{\bf Proof}:

By using local convergence, one can easily  show that there exists a fixed $\epsilon>0$, such
that for any non-trivial bubble ${\tilde u}_b$, $E_\lambda({\tilde
u}_b)>\epsilon.$ Same conclusion holds for
non-trivial principal component ${\tilde u}_{p}:S^1\times {\bf
R}^1 \setminus \{ \mbox double\, points\}\rightarrow{\tilde M}$.
Note that here we used $x_{-}\not =x_{+}.$

Since 
\begin{eqnarray*}
E_\lambda({\tilde u}) & =& \Sigma_{b, p}E_\lambda({\tilde
u}_b)+E_\lambda({\tilde u}_p)\\
& = &\int_{S^1}x_+^*\lambda-\int_{S^1}x_-^*\lambda=c_+-c_-
\end{eqnarray*}
is fixed, we only need to prove that the number of trivial principal
components and trivial bubble components is uniformly bounded. 

 Now for each trivial principal component,
there exists some  non-trivial bubble components lying on the same target component.
 Therefore, the
number of such components is less than or equal to the number of non-trivial bubble
components, which is bounded. Finally, it is easy to see inductively that the number of
trivial bubble components can be uniformly bounded by the number of non-trivial
bubbles and principal components. 

\begin{lemma}\label{lemma6}

Given ${\tilde u}\in [{\tilde u}]\in {\overline{\cal M}}(x_{-}, x_+; J),$
if $x$ is a closed orbit such that it is an intermediate end of some
component of ${\tilde u}$. Then $\int x^*\lambda <\int x_+^*\lambda=c_+$.
Therefore, there are only finite such closed orbits. 
\end{lemma}

{\bf Proof}:

$x_+\subset{M}_{p, +}$ is the only closed orbit lying
on the positive end of ${\tilde M}_p$, where ${\tilde M}_p$ is the most
right (positive) components of the target of ${\tilde u}$. Then
$$\int x_+^*\lambda-\Sigma_i \int x_{-, i}^*\lambda=\int_{u|{{\tilde
M}_p}}d\lambda >0, $$ 
where $x_{-, i}$ is one of the closed orbits on ${\tilde M}_{p, -}$
appeared as a non-trivial end of ${\tilde u}$. The conclusion follows by
induction.

To see that there are only finite such intermediate $x$, we use Remark
3.2. It follows from the remark there that $\int x^*\lambda$ is bounded
below by $\epsilon>0$ of the lower bound of the $E_{\lambda}$-energy
of non-trival bubbles. 

\QED

It follows from this argument that the number of double points of a
component of a stable map appeared in the compactification is also
bounded.

Because of these lemma, the proof of the theorem essentially can be reduced to the case
that $[{\tilde u}_i]^\infty_{i=1}$ has only one component and we need to
show that such a sequence has a weak-limit $[{\tilde u}_\infty]\in
{\widetilde{\cal M}}(x_-, x_+; J).$ 

Note that the above two lemmas together implies that the energy of
$E({\tilde u})$ is uniformly bounded for any ${\tilde u}\in 
{\widetilde{\cal M}}(x_-, x_+; J).$ 

$\bullet$ {\bf Stabilization of the target and its local deformation}

The target of ${\tilde u}$ of a stable map is a union $U=\cup_{i\in
I_P}{\tilde M}_{p_i}$. Each ${\tilde M}_{p_i}\simeq {\tilde M}=M\times
{\bf R}$ with a ${\bf R}^1$-symmetry coming from the r-translations
along the second factor ${\bf R}$. To stabilize $U$, we add a marked
point $z_i$ on the second factor of ${\tilde M}_{p_i}$ to remove theses
symmetries.  Given $({\tilde M}_i, z_i)=M\times ({\bf R}, z_i)$,
$i=1, 2$ and $\tau\in {\bf R}^+$ of  a deformation (gluing) parameter,
we can form $U_\tau=({\tilde M}_1, z_1)\#_\tau ({\tilde M}_2, z_2)$ of
the local deformation of $U=({\tilde M}_1, z_1)\cup ({\tilde M}_2, z_2)$
with respect to the parameter $\tau$ by the obvious gluing construction
along the second factor of ${\tilde M}_1$ and ${\tilde M}_2$, namely,
cutting off $r_1>\frac{1}{\tau}$ of ${\tilde M}_1$ and
$r_2<-\frac{1}{\tau}$ and gluing back the remaining parts. Then
$U_\tau\simeq M\times {\bf R}$ with two marked points $z_1$ and $z_2$ on
${\bf R}$. Similarly, if $U=\cup_{i\in I_P}{\tilde M}_{0_i}$ and
$\tau=(\tau_1, \cdots, \tau_{\gamma-1})\in ({\bf R}^+)^{\gamma-1}$, 
$\gamma=\#(I_P)$, with $\gamma$ marked points $z_i\in {\tilde M}_{p_i}$,
$i=1,\cdots, \gamma,$ we can form $U_\tau={\tilde
M}_{p_1}\#_{\tau_1}{\tilde M}_{p_2}\#\cdots\#_{\tau_{\gamma-1}}{\tilde
M}_{p_\gamma}$ with marked points $z_1, \cdots, z_\gamma$ on it. 
Another way to think this is to treat the marked point $z_{i}$ as a
marked section $M\times \{z_{i}\}$ in ${\widetilde M}_{i}$.

A cylinder ${\tilde M}=M\times ({\bf R}; -\infty, +\infty, z_1, \cdots,
z_n)$ with two end points $-\infty$ and $+\infty$ and $n$ distinct
marked points $z_1, \cdots, z_n$ is said to be stable of type $M$ if
$n\geq 1$. let ${\cal M}_n(M)$ be the collection of all such stable
cylinders of $n$ marked points of type $M$. Then it has an obvious
compactification $${\bar{\cal M}}_n(M)={\cal
M}_n(M)\coprod_{\tiny\begin{array}{l}
l+m=n\\
l, m\geq 1
\end{array}}{\cal M}_l(M)\times {\cal M}_m(M)).$$ The topology of
${\bar{\cal M}}_n(M)$ near the boundary points is described by the local
deformation (gluing) above.

\vspace{3mm}

$\bullet$ {\bf Weak Convergence}

$\bullet$ Stabilization of a semi-stable curve and its local deformation:

Domain $\Sigma$ of a stable map ${\tilde u}$ is only a semi-stable
curve. Therefore, there may exist some non-trivial bubble components or
principal components whose domains contain only one or two double
points. We can stabilize these unstable components by adding minimal
number of marked points ${\underline y}=(y_1, \cdots, y_m)$ to get a
stable curve $(\Sigma, {\underline y}).$
In particular, for each top (hence, unstable ) bubble, the symmetry
group is three dimensional because of extra structure of the 
$S^1$-parametrization at infinity along its end. To stabilize such a component
we introduce an arbitary marked point $y_{1}$ first. Then the $S^1$-parametrization
at infinity together with the marked point determine a marked ray connecting
$y_{1}$ to 
$\theta =0 $ at  $S^1$ at infinity 
an obvious way. We add the second marked point $y_{2}$ on the marked ray
with distance of $1$ to $y_{1}$ to get the desired stablization.
 Let $(\Sigma_{\alpha},
{\underline y})$ the local deformation of $(\Sigma, {\underline y})$ in
the moduli space of stable curves, where $\alpha$ is the collection of
deformation parameters associated with double points of $\Sigma.$ Note
that the moduli space of stable maps used here is not the ususal
Degline-Mumford compactification but an obvious modification of the
moduli space of stable $(J, H)$-maps used in [LT]. Here for each ordinary
double point of ${\tilde u}$, we associate it with a complex gluing
parameter and for each double point corresponding to an end approaching to
closed orbit, we associate a positive real gluing parameter. 

To see this more concretely, we consider the following example.

{\bf Example}

Consider the semi-stable curve $\Sigma= P\cup_{d_{1}=d_{2}}B$
with principal component $P={\bf R}\times S^1$ and bubble $B$ joint
at the double point $d$. Assume that the double point corresponds to the
two end on $P$ and $B$. Not that on $P$ there are other two marked
points corresponding to $-{\infty}$ and $ +{\infty}$ in
 ${\bf R}\times S^1$. The moduli space of such semi-stable
map is $1$-demensional due to the choices of $S^1$-parametrization at
$d$. To stable such a $\Sigma $,  we only need to stabilize $B$, which
is described above. Associated to the double point, there is a 
one dimensional local deformation of $\Sigma$ with respect to a gluing parameter 
$\alpha \in {\bf R}^+$. By letting $\Sigma$ vary in above $1$-dimension 
moduli space, the deformation gives the elements in the moduli
space ${\cal M}_{0, 4}$, which form a neighbourhood the
the $1$-dimensional moduli space.  Therefore, the above $1$-dimensional
 moduli space can be thought as part of the boundary of
 ${\cal M}_{0,4}$. Of course, this not the usual Degline-Mumford
compactification of ${\cal M}_{0,4}$. 
On the other hand,  give any sequence $\Sigma_{i}\in  {\cal M}_{0,4}$
with $\Sigma_{i} = ({\bf R}\times S^1; -\infty, +\infty, y_{i}, z_{i}$ with
$y_{i}\rightarrow z_{i}$ as in the bubbling below, the process there will
give a limit of the sequence in the above $1$-dimensional moduli space.

$\bullet$ {\bf Defintion of Weak Convergence}:

Given $[{\tilde u}_i]^\infty_{i=1}\in {\overline{\cal M}}(x_-, x_+;
{\tilde J})$, we say that $[{\tilde u}]$ is weakly $C^\infty$-convergent to
$[{\tilde u}_\infty]\in{\overline{\cal M}}(x_-, x_+; {\tilde J})$ if there
exist ${\tilde u}_i\in[{\tilde u}_i]$ and ${\tilde u}_\infty\in[{\tilde
u}_\infty]$ such that 

(i) After adding some marked points ${\underline y}_i$ to $\Sigma_i$,  $(\Sigma_i,
{\underline y}_i)$ is convergent to the minimal stabilization of
$(\Sigma_\infty, {\underline y}_\infty)$ in the moduli space of stable
curves, where $\Sigma_i$ and $\Sigma_\infty$ are the domains of ${\tilde
u}_i$ and ${\tilde u}_\infty$. 
Note that the number of marked points ${\underline y}_i$'s is same as the number of
marked points ${\underline y}_\infty.$ Therefore, when $i$ is large enough,
there exists an $\alpha_i$ such that $(\Sigma_i, {\underline y}_i)$ is
equivalent to $(\Sigma_{\infty, \alpha_i}, {\underline y}_\infty)$ of the
deformation of $(\Sigma_\infty, {\underline y}_\infty)$ with respect to
the gluing parameter $\alpha_i$. 

 Let $\phi_i:(\Sigma_{\infty,
\alpha_i}, {\underline y}_\infty)\rightarrow(\Sigma_i, {\underline
y}_i)$ be the equivalence map. 

(ii)Let $U_i=\cup_{j\in I_{P_i}}{\tilde M}_{p_{i,j}}$ and $U_\infty=
\cup_{j\in I_{P_\infty}}{\tilde M}_{{p_\infty}}, $ be the targets of 
${\tilde u}_i\in[{\tilde u}_i]^\infty_{i=1}$ and ${\tilde
u}_\infty\in[{\tilde u}_\infty]$.
We stabilize $U_\infty$ by adding minimal number of marked points
${\underline z}_\infty$ and require that after adding same number of
makred points ${\underline z}_i$ to $U_i$, $(U_i, {\underline z}_i)$ is
convergent to $(U_\infty, {\underline z}_\infty)$ in the space of
${\bar{\cal M}}_n(M)$. Here $n$ is the number of marked points of
${\underline z}_\infty$. Therefore, there exists $\tau_i$ such that
$(U_i, {\underline z}_i)$ is equivalent to
$((U_\infty)_{\tau_i},{\underline z}_\infty)$. 

Let $\psi_i:(U_i, {\underline z}_i)\rightarrow ((U_\infty)_{\tau_i},
{\underline z}_\infty)$ be the equivalence map. Note that for $u_i$
closed to $u_\infty$, the gluing parameter $\alpha_i$ of the domain 
$(\Sigma_i, {\underline y}_i)\equiv((\Sigma_\infty)_{\alpha_i},
{\underline y}_\infty)$ is not compleletely independent of the gluing
paprameter $\tau_i$ of the targe $((U_\infty)_{\tau_i}, {\underline
z}_\infty)$ since along these ends where $[u_\infty]$ approaches closed
orbit $\alpha_i=0\Longleftrightarrow \tau_i=0.$ However, 
when $\alpha_i \not =0$,
 hence $\tau_i\not =0$, they are essentially independent each other.

(iii) Given a compact set $K\subset\Sigma_\infty\setminus\{\mbox{double
points}\}$, the compact image ${\tilde u}_\infty(K)\subset
U_\infty\setminus\{\mbox{end of $U_\infty$}\}$. Hence for $i$ large enough,
${\tilde u}_\infty(K)\subset(U_\infty)_{\tau_i}\equiv U_i$. Therefore, $\psi^{-1}_{
i}\circ u_\infty$ is well-defined on $K$ and  it maps $K$ into $U_i$. On
the other hand, for large $i$, $K\subset(\Sigma_\infty)_{\alpha_i}$ and
$\phi_i(K)\subset\Sigma_i$, and ${\tilde u}_i\circ\phi_i:K\rightarrow U_i$. We
require that (a) ${\tilde u}_i\circ\phi_i|_K$ is $C^0$-close to $(\psi^{-
1}_{i}\circ {\tilde u}_\infty)|_K$ when $i$ is large enough, hence, 
$\psi_{i}\circ u_i\circ\phi_i|_{K}:K\rightarrow U_\infty $ is well-defined.
(b)  for any compact subset $K\subset
\Sigma_\infty\setminus\{\mbox{double points}\}$, 
 $\psi_{i}\circ {\tilde u}_i\circ\phi_i|_{K}$ is
$C^\infty$-convergent to ${\tilde u}_{\infty}|_{K}$.

Note that $E_\lambda({\tilde u}_i)=E_\lambda({\tilde u}_\infty)=c_+-c_-$
is fixed. This  together with the two statements of Sec. 4. imply 
that the projections of the images
of ${\tilde u}_i$ to the contact manifold $M$ is $C^{0}$-close to the projection of the image
${\tilde u}_\infty$, and 
that near a closed orbit $x$ with $\lambda$-period $c$
as an asymptotic
end of ${\tilde u}_\infty$, along the non-compact ${\bf R}$-direction of ${\tilde M}$,
${\tilde u}_i$ is essential same as the function $c\cdot s$,
  when $i$ is large enough. 

We start with a detailed description on the case that the sequence
$[{\tilde u}_i]$ only develops one bubble, as  it  already exhibits
all of the main points of the general case. The following lemma plays an
important role both in the proof of this theorem and in the later formal
dimension counting of the boundary of the modui space of ${\tilde J}$-
holomorphic maps. 

\begin{lemma} If $\{[{\tilde u}_i]\}^\infty_{i=0}\in {\cal M}(x_-, x_+;
{\tilde J})$ develops only one bubble at its limit $[u_\infty]$, then
the target $U_\infty$ contains at least two elements and the image of
bubble is not in the right most component. This implies that the domain
$\Sigma_\infty$ of $u_\infty$ contains at least three components. 
\end{lemma}

\noindent {\bf Proof}:

Let ${\tilde u}_i\in [{\tilde u}_i]$, ${\tilde u}_i=(u_i,
a_i):\Sigma_i={\bf R}^1\times S^1\rightarrow {\widetilde M}$, where
$u_i:{\bf R}^1\times S^1\rightarrow M$ and 
$a_i:{\bf R}^1\times S^1\rightarrow {\bf R}$. By assumption,
there exists bubble point $y_i\in \Sigma_i$ such that $|d{\tilde
u}_i(y_i)|\rightarrow\infty$. First assume that $y_i$ stays in a compact
set of $\Sigma_i={\bf R}^1\times S^1$, hence $y_i\rightarrow y_\infty
\in {\bf R}^1\times S^1$ as $i\rightarrow \infty$ after taking a
subsequence. We claim that $|a_i(y_i)|$ is not bounded and $a_i(y_i)$
tends to $-\infty.$ Otherwise, assume that $|a_i(y_i)|<C$. Then
$a_i(y_i)\rightarrow a_{\infty, y}\in {\bf R}.$ Now the domain of
$\Sigma_i$ has two marked points $y_i $ and $w_i$, where $w_i$ is a
point on the circle centered at $y_i$ of radius $\frac{1}{|d{\tilde
u}_i(y_i)|}$. Note that here we can make an arbitary choice for $w_{i}$ on the 
circle. See the remark after on how  to make "correct" choice.
These two marked points $y_i, w_i$ together with $-\infty,
+\infty$ on $S^2=\Sigma_i\cup\{-\infty, +\infty\}$ have moduli, and we
can 
identify  the domain $(\Sigma_i, -\infty, +\infty, y_i, w_i)$ with 
$(\Sigma_i, -\infty, \infty, d)\#_{\alpha_i}(S^2, 0, 1, d')$. 
Here $(\Sigma_i;
-\infty, \infty, d)\#_{\alpha_i}(S^2; 0, 1, d')$ is obtained from
$(\Sigma_i; -\infty, \infty, d)\bigvee_{d=d'}(S^2; 0, 1, d')$
by gluing at $d$ with some deformation parameter $\alpha_i\in {\bf
R}^+$ with $\alpha_i\rightarrow 0$ as $i\rightarrow\infty,$  and 
$(\Sigma_i; -\infty, \infty, d)\bigvee_{d=d'}(S^2; 0, 1, d')$
is one of the elements in the $1$-dimensional moduli space mentioned in
the previous  example. In particular, along the end $d$, there is a 
$S^1$-parametrization.  Intuitively, what we did here is to 

conformally enlarg  a small disc of $\Sigma_i$ near $y_i$, bringing
$y_i$, $w_i$ into standard points $0, 1$ in standard disc. 
 
Now  $(\Sigma_i\#_{\alpha_i} S^2; -\infty, \infty, 0 ,1)$ 
 has four
marked points $-\infty, \infty, 0, 1$, and 
$$(\Sigma_i\#_{\alpha_i} S^2; -\infty, \infty, 0 ,1)\simeq (\Sigma_i;-
\infty, \infty, y_i, w_i).$$
Let $\phi_i:(\Sigma_i\#_{\alpha_i} S^2; -\infty, \infty, 0,
1)\rightarrow(\Sigma_i;-\infty, \infty, y_i, w_i)$ be the identification
map. 
Let $D_R$ be the half shpere glued with a finite cylinder
$S^1\times [0; { R}]$ along its boundary. We still use $D_R$ to
denote its obvious conformal image in $(S^2; 0, 1, d')\subset 
(\Sigma_i; -\infty, \infty, d)\bigvee_{d=d'}(S^2; 0, 1, d')$ centered at
$0$, and $D_{R, i}$ the corresponding image in $\Sigma_i\#_{\alpha_i}S^2$
when $i$ is large enough. 

Define ${\widetilde V}_{i, R}=({\tilde u}_i\circ\phi_i)|_{D_{R, i}}$.
Then as we did before for bubbling, ${\tilde v}_{i,
R}\rightarrow{\tilde v}_{\infty, R}:D_R\rightarrow{\widetilde M}$ and
${\tilde v}_i={\tilde u}_i\circ\phi_i$ is locally $C^\infty$-convergent
to ${\tilde v}_\infty=\cup_R{\tilde v}_{\infty, R}:D_\infty=D^2\cup({\bf
R}^+\times S^1)\rightarrow{\widetilde M}$. That is $\{{\tilde u}_i\}$
produce a bubble at $y_i$. The domain of ${\tilde v}_\infty$ is the
complex plane but thought as half sphere with a half infinite cylinder
attached. Since ${\tilde v}_\infty$ is ${\tilde J}$-holomorphic and
$E({\tilde v}_\infty)<\infty$, $|D{\tilde v}_\infty|$ is uniformly
bounded. As before, $\lim_{s\rightarrow +\infty}{
v}_{\infty}(s, t)=x(t)$ of some periodic orbit along its cylindrical
end. Now fix $\epsilon>0$, and consider ${\tilde u}_{i,
\epsilon}={\tilde u}_i|_{\Sigma_i\setminus D_\epsilon(y_i)}.$ By our
assumption that there is only one bubble we conclude that for any fixed
$\epsilon>0$, $|d{\tilde u}_{i, \epsilon}|<C_\epsilon $ for any $i. $ We
may assume that $\lim_i{\tilde u}_i(0, 0)$ exists at the begining and
$y_i\not = (0, 0). $ Then the same argument as before implies that ${\tilde u}_{i,
\epsilon}$ is $C^\infty$-convergent to ${\tilde u}_{\infty,
\epsilon}:{\bf R}^1\times S^1\setminus
D_\epsilon(y_\infty)\rightarrow{\widetilde M}.$ Here we used that
$|y_i|$ is bounded and hence $y_i\rightarrow y_\infty\in {\bf R}^1\times
S^1.$ By letting $\epsilon\rightarrow 0$, we get ${\tilde
u}_i|_{\Sigma_i\setminus \{y_i\}}$ is locally $C^\infty$-convergent to
${\tilde u}_\infty|_{{\bf R}^1\times S^1\setminus \{y_\infty\}}.$
 Identifying $D_{\epsilon}(y_\infty)-\{y_\infty\} $ ($\subset {\bf R}^1\times S^1-
\{y_i\}$) with ${\bf R}^+\times S^1, $ then
$\lim_{s\rightarrow\infty}{ u}_\infty(s, t)=x'(t)$ of a closed
orbit. 

Let ${\tilde v}_\infty=(v_\infty, b_\infty)$. Then
$\lim_{s\rightarrow\infty}b_\infty(s, t)=+\infty$. Otherwise, since
$b_\infty(s, t)\sim cs+d$ with $c=\int_{S^1}x^*\lambda \not = 0$, we
have $\lim_{s\rightarrow\infty}b_\infty(s, t)\rightarrow -\infty.$ But
since $\Delta b_\infty \geq 0$, this contradicts to the maximal
principle for sub-harmonic functions. 

Therefore, $b_\infty(s, t)\sim cs+d$ with $c=\int_{S^1}x^*\lambda >0.$
The induced orientation of ${\tilde v}_\infty$ on $x$ is the same as the
one given by $\lambda.$ By our assumption that there is only one bubble,
if we set ${\tilde u}_\infty=(u_\infty, a_\infty)$, then
$\lim_{s\rightarrow \infty}a_\infty(s, t)=+\infty$,
$\lim_{s\rightarrow\infty}{ u}_\infty(s, t)=x'(t)=x(t).$ Here,
$(s,t)\in {\bf R}^+\times S^1=D_{\epsilon}(y_{\infty})\setminus \{y_{\infty}\}$.

 This implies that the induced orientation on
$x'(t)=x(t)$ form ${\tilde u}_\infty$ is also the same as the one given
by $\lambda.$ However, ${\tilde u}_\infty\cup {\tilde v}_\infty$ is the
weak limit of ${\tilde u}_i$ and $x(t)=x'(t)$ is the limit of some
corresponding curves $x_i$ in ${\tilde u}_i$. Clearly, the induced
orientations of $x_i$ obtained from the two sides of ${\tilde u}_i$ are
opposite to each other. This is a contradiction. 

We remark that one can also get an alternative proof of above statement
by using gluing in [LT] and maximal principle instead of using this
orientation consideration. 

This proves that $|a_i(y_i)|$ is not bounded under the assumption that
$|y_i|$ is bounded. In the case that $y_i\rightarrow \pm\infty$, 
$({\bf R}^1\times S^1; y_i, 0, -\infty, \infty)$ tends to a boundary
point of moduli space ${\bar{\cal M}}_{0, 4}$. If, say,
$y_i\rightarrow\infty$, let $(S^2; -\infty, y_\infty,
d)\bigvee_{d=d'}(S^2, d', 0, +\infty)$ be the limit curve. Then $({\bf
R}^1\times S^1; -\infty, y_i, 0, \infty)\simeq (S^2\#_{\alpha_i}S^2; -
\infty, y_\infty, 0, \infty)$ for some $\alpha_i\in {\bf C}^*$. Now in
$S^2\#_{\alpha_i}S^2$, $y_\infty$ plays the same role $y_i$ in
$\Sigma_i$ but it stays away from the two ends. The above argument is
still applicable except that at the limit, the domain has one more splitting.

Therefore, $a_i(y_i)\rightarrow\pm\infty$. If $a_i(y_i)\rightarrow
+\infty$, after shifting by $-a_i(y_i)$ to the target of ${\tilde u}_i$
and define ${\tilde w}_i=(u_i, a_i-a_i(y_i))$, the above argument is
still applicable to ${\tilde w}_i,$ and we get bubble at $y_{\infty}$, still
denoted by ${\tilde v}_{\infty}=(v_{\infty}, b_{\infty})$. In particular,
$\lim_{s\rightarrow\infty}b_{\infty}(s,t)=+\infty.$ 
Therefore, we get a bubble as before but with target ${\widetilde M}'$
lying on the right of ${\widetilde M}$ with the end of the bubble
approaching a closed orbit lying on the right end of ${\widetilde M}'$.
As before, the orientation consideration and maximum principal
 rule out this possibility.

Therefore, $a_i(y_i)\rightarrow -\infty$.  Of course, we still can
define ${\tilde w}_i$ by the same formula above. Arguing as before,
we conclude that we still get a bubble from $\{{\tilde
w}_i\}^\infty_{i=1}$, still denoted by ${\tilde v}_\infty$,
$D_\infty=D^2\cup({\bf R}^1\times S^1)\rightarrow{\widetilde M}'$. But
the target ${\widetilde M}'$ lying on the left end of ${\widetilde M}$,
and $\lim_{s\rightarrow\infty}{v}_\infty(s, t)=x(t)$ in the
right end of ${\widetilde M}'$. For simplicity, assume that there is no
further splitting of the target. (This follows form our assumption that
there is  only one bubble if we also count "connecting bubbles".) Then
as before, ${\tilde w}_i|_{\Sigma_i-\{y_i\}}$ is locally $C^\infty$-
convergent to ${\tilde w}_\infty|_{{\bf R}^1\times S^1-\{y_\infty\}}$ (again
assume first that $|y_i|<c$ and use deformation as before to deal with
general case), and along the end $D(y_\infty)-\{y_\infty\}
\simeq{\bf R}^1\times  S^1$, 
$$\lim_{s\rightarrow\infty}{\tilde w}_\infty(s,
t)=x'(t)=x(t)\in{\widetilde M}_-={\widetilde M}'_+.$$

To see that there is at least one more component of the domain in the
limit of $[{\tilde u}_i]$, we note that each $u_i$ connects $x_-\in
{\widetilde M}_-$ to $x_+\in {\widetilde M}_+$, therefore, there exists
$s_i$ such that ${\tilde u}_i|_{(-\infty, s_i]\times S^1}$ lies on the
"left" of ${\tilde u}_i(y_i)$. Now ${\tilde u}_{i, -R}={\tilde u}_i|_{(s_i-
R, s_i)\times S^1}$ is $C^\infty$-convergent to ${\tilde
u}_{\infty, -R}$ after identifying $(s_i-R, s_i)\times S^1$ with $(-R,
0)\times S^1. $ We get ${\tilde u}_{\infty, -\infty}=\cup_R{\tilde
u}_{\infty, -R}:(-\infty, 0)\times S^1\rightarrow{\widetilde M}'$. We
only need to show that for some $R$, ${\tilde u}_{\infty, -R}$ is not
constant map. However, since ${\tilde u}_i$ asymptotically approximates 
to $x_{-}$ with exponential decay as $ s$ tends to $-\infty$. More precisely,
we have $|a_{i}(s,t)-(cs+d_{i})|< e^{-k_{i}s}$ for some $k_{i} > 0$. Note that
$c=-\int_{S^1} x^*_{-} \lambda \not = 0$ is the same for all i. Therefore 
we can replace ${\tilde u}_{1}(s,t) $ by ${\tilde u}_{i}(s+s_{i}, t)$
 with some very
negative $s_{i}$ such that $|d{\tilde u_i}|_{[0,R]\times S^1}| >\epsilon>0 $ 
for some fixed $\epsilon$.  Then the above limit 
${\tilde u}_{\infty, -\infty}$ is not constant.

This proves the lemma. Note that the component
${\tilde u}_{\infty, -\infty}$ of the limit  could come from 
a closed orbit, i.e. it is 
a trivial principal components. However, in this case, there is a bubble
component lying in the same component of the target. 

We remark that in the general case with multi-bubbling, the same proof above 
proves that each of  bubbles lie on some new component of targets which lie on the 
left of the original ${\tilde M}$. Moreover, there is at least one 
more principal
component lying on the the new "left" component. In particular, in the
"new" component of the target, where the first top bubble lies on, there
are at least two connected components of the domain of the limit.

\QED

\begin{remark}
Some remark on the special role played by the marked point $(y_{i},
w_{i})$ in the above lemma and some related issue is in order. 
Recall that $y_{i}$ is the
 point where $ |d u(y_{i})|\rightarrow \infty$ and $w_{i}$ is the
 point lying on the cirle of radius $\frac {1} {|d u(y_{i})|}$ 
measured in the standard metric on ${\bf R}\times S^1$. In the process of
 bubbling we bring $(y_{i}, w_{i})$ into the standard point $(0,1)$ in
$S^2$. On one hand, the point $w_{i}$ will be used to determine the side
of bubbling at each stage, on the other, it will also determine two marked
lines on the two ends of $\Sigma_{\infty}$ joint at the double point of 
$\Sigma_{\infty}$. Since the two components $v_{\infty}$ and $u_{\infty}$ of
the limit approach to a closed orbit $\{x\}$, these two marked lines will
specify the base point $1\in S^1$ and hence we get a particular paremetrized
$x: S^1\rightarrow M$ in $\{x\}$.  However, in the bubbling one can make
arbitary choices for $w_{i}$ the the cirle. This implies that in the
 compactification below, we can use fixed parametrized closed orbits
as asymptotic limit to which bubble components, and hence the adjecent principal
components, approach along their parametrized ends. On the otherhand,
the different $S^1$-paramerizations associated to each of such ends contributs
 an one dimensional moduli to the domain of the limit stable map. 
An equivalent way to think about this is to fix an  $S^1$-parametrization 
for each of
such ends of the limit curve. Then we can not  fix the parametrizations
of limit closed orbits anymore. 

Note that in the case of  splitting of principal components, as
the maked lines are already fixed a priori, clearly, 
all elements in $\{x \}$ may
appear in the limit.

\end{remark}

\begin{remark}

The proof of this lemma can be used the deduce some furth
restrictions on the possible domains of stable maps, which
appeared in the compactifiction of ${\cal M}(x_{-}, x_{+}, J)$
(meaning as a limit of some sequence of elements
in ${\cal M}(x_{-}, x_{+}, J)$).
There are two general requirements. The first one is that 
the maximum principal for the $a$-component of the stable map
must hold
(as well as the closed related orientation consideration should
be incorperated). The second is that there is no loop in the set
of components of a domain. Applying these two requirements to the 
case that there are two connected components of a stable map
lying same component of the target with one ordinary double point
joint the two components of the domain, one conclude that each
connected component has at least one end lying on the positive end
of the component of the target. Starting from this, inductively one
can prove that in the rightmost (positive) component of the target,
there are at least two closed orbits on the positive end of the
component of the target, which appeared as the asymptotic limits
of the stable map. However, the orintation consideration as in the 
proof of the last lemma ( or the maximum principal plus gluing),
implies that this is impossible.

Therefore, we conclude that all double points of a stable map
in the compactification are ends. The same consideration also implies
the following simple picture one the structure of the components of a
stable map appeared in the compactification. Starting from the
leftmost  component
whose "left" asympototic end is $x_{-}$, there exist one and only one end
of this component, along which the component approaches to a closed orbit
$x_{1}$
on the positive end $M_{+,1}$ of component of the target. It is easy to see
that all the other ends of the components must lie on the negative end of the
component of the target. In this case, since we are already in the leftmost
 component, this is impossible. However, this can happen in general case
and we will use this to do induction in a moment.  If the next adjecent
component lying on the adjecent component of the target, we are in the
same position as before and we can inductively go further. We now show that
this must be the case. Otherwise, the new component still stay in the
 same component of the target, then the induced orientation on $x_{1}$ from
the two adjecent components are the same, which is a contradiction.
 We conclude that there is a chain of components
( should be called principal components ), each lying on different but
adjecent components of the target and each connecting two closed
orbits on the two different ends of the component of the target.
As mentioned above, for each of the principal components, all the other
ends (if there are any) must lie on the negative end of the component
of the target by maximum principle and gluing. 

To get a complete picture, we need to know the behavior of  those adjecent components
to those negative ends of, say,  a typical  principal component. 
The orientation consideration implies that each of such components
must lie in the left adjecnt component to the component of the target, on which
the principal component lies. The maximum principle and gluing implies 
that the end at which the principal component and the new adjecent
component joint together is the only positive end for the new component. 
Now we are in the position of induction and we get a very simple
structure on the components of a stable map which appears as a limit map.
Namely, each component of a stable limit connecting map has only one positive
end and possibly many negative ends without any ordinary double points.
Starting form the (only) rightmost end $x_{+}$, all components of the stable map
form a tree pointed to negative $a$-direction.

It follows from this that for each intermediate closed orbit $x$, which
appears as an end of the limit stable map connecting $x_{-}$ and $x_{+}$,
$\int x^*\lambda$ is bounded above by $\int x^*_{+}\lambda$ and bounded
below by the minimum of  $\int x^*_{-}\lambda$ and $\epsilon$, the
lower bound of the $E_{\lambda}$-energy of non-trival bubbles.
This is used in the proof of Lemma 3.3.

%the maximum principle (and gluing)  implies that there is another end along which
%the component approaches to a closed orbit $x_{1, 2}$ still on
%$M_{+,1}$. The no loop requirement implies that there is only one such
%end. In other words, all other ends of the component will go to the
%negative end of the component of the target. 

%Now  we can study more on what happen to these 
%ends which  go to the negative end. Starting from each of such end,
%there is a tree of finite components, each of components will produce at
%least one more closed
% there is a chain of components
%( should be called principal components ), each lying on different but
%adjecent components of the target and each connecting two closed
%orbits on the two different ends of the component of the targethaving two 

\end{remark}

To prove the compactness in general, as in the usual Gromov-Witten theory
or Floer homology, there are three steps (i) formation of all bubbles which
lie on the top of the bubble tree; (ii) local convergence of the sequence of 
$\{ {\tilde u}_{i}\}^{\infty}_{i=0}$ along the base, including
splitting or degeneration of  principal components; (iii) formation of 
the intermediate bubbles and related "zero bubbling" along connecting
necks. Most of analytic part of  the proof for these are the analogy to 
the symplectic case, except the two statments concerning the exponential
decay
of a bubble along its non-removable singularity and and the behavior of 
"connecting neck" along the non-compact ${\bf R}$-direction, detailed in
 Sec.4. We will only outline the those parts whose
proof are similar to the symplectic case.

To do the step (i), we proceed inductively as in the usual symplectic case.
The proof of the above lemma serves as the staring point of the induction.
During the formtion of the first bubble, the domain of 
${\tilde u}_{i}$ is deformed into $(\Sigma_i; y_i^1, w_i^1)$, where
  $y_i^1, w_i^1$ are the maked points denoted by $y_i, w_i$ in the previous
lemma. But we think $\Sigma_i$ as ${\bf R}^1\times S^1$ with a small disc
centered at $y_{i}$
removed, then gluing back a portion of a cylinder, $[0, R_{i}]
\times S^1$ with a half sphere attached. In this model of 
$(\Sigma_i; y_i^1, w_i^1)$, the maked points $y_i^1, w_i^1$ becomes the
 standard points $0, 1$ in the half sphere. Here $R_{i}=\frac{1}{\alpha}$
 and $\alpha_{i}$ is the deformation parameter in the Lemma
before. The target ${\tilde M}$ originally has three marked sections
$-\infty, +\infty, 0$. We introduce a new marked section $z^1_{i}=a_{i}(y_{i})$, 
where $a_{i}$ is the second factor of ${\tilde u}_{i}$. As proved above,
$z^1_{i}<0$ and $|z^1_{i} - 0 | \rightarrow\infty$ as $i \rightarrow \infty$.

We then check that if $|d {\tilde u}_{i}|$ measured in the induced metric 
in the new deformed domain is uniformly bounded. Assume that is it not. Since
the injective radius of these new domains are bounded below, we can repeat the 
process before to produce second bubble  by introducing new marked points
$y_i^2, w_i^2$  in the domain and marked section $z^2_i$ in the  taget
which play the same role as $y_i^1, w_i^1$ and $ z^2_{i}$  in the formation of
the first bubble.  
%Note that in the case that 
%the new bubbling point $y_{2}$ lying on the first bubble, we have
% $z_i^2 < z_i^1 $ and $|z_i^2-z_i^1|\rightarrow 
%\infty$. 
As each bubble has a minimal amount of $E_{\lambda}$-energy
bounded below, this process will stop after finite steps. We end
up
with a deformed new domain $(\Sigma_i; y_i^1, w_i^1, y_i^2, w_i^2,\cdots 
y_i^k, w_i^k)$ of ${\tilde u}_{i}$. As above, we think it 
 as ${\bf R}^1\times S^1$ with $k$-small disc
centered at $y_{i}^j, j=1,\cdots\, k$
removed, then gluing back a portion of a cylinder, $[0, R_{i}^j]
\times S^1$ with a half sphere attached.
As before the maked points $y_i^j, w_i^j$ in 
$(\Sigma_i; y_i^j, w_i^j; j=1,\cdots, k)$,  becomes the
 standard points $0, 1$ in these $k$-half spheres. The target ${\tilde M}$
of ${\tilde u}_i$
now has marked points $-\infty, +\infty, 0, z_i^j, j=1,\cdots,k.$
%As above in the  case that each $y_{i}^j$ lying on the previous
%bubble , we have  
% that $ z_i^{j+1}<z_i^j$ and $|z_i^{j+1}-z_i^j|\rightarrow \infty$.

Now 
$|d {\tilde u}_{i}|$ measured in the induced metric 
in the new deformed domain is uniformly bounded. Let $D_{i, R}^j, j=1,\cdots,
k,$ be one of the $k$ half spheres centered at $0=y_{i}^j$ with a portion of a
cylinder of length $R$ attached in the deformed domain $\Sigma_i$ and 
$D_{i, R}$ be their union.  We will use $B_{i, R}$ to denote 
 the subset of $\Sigma_{i}$
obtained by removing a small disc around each of those $y_i^j$ which produces
a "top" bubble,  and then gluing back a cylinder of length $R$. Then for any
 fixed $R$, ${\tilde u}_{i}|_{D_{i, R}}$ is 
$C^{\infty}$-convergent. By letting $R\rightarrow \infty$, we  obtained all
top bubbles. 

On the other hand, by restricting ${\tilde u}_{i}$
to part of $ {B_{i, R}}$  of, say, length $R$ and shiftng the target with
a suitable constant, we get   the local convergence along the "base" after 
letting $R$ tend to infinity. Note that in the local convergence of the base,
the domain may splitting further into broken connecting maps. It is possible
that only one of the two ends of some component of such a broken connecting
map approaches to a cloed orbit, the other is just a double point.
Note  also that during the process of these local convergenes  and
bubbling, the target also
gets split into severl components. For example in the case that each of
the distances between these $k$ marked 
sections $z_{i}^j, j=1,\cdots, k$  tends to infinity, the target of the limit
has at least $k+1$ components.
This essentially finishes the first two steps (i) and (ii).

It may happen that for some
of $B_{i,R}$, the limt of the local convergence is only a constant map.
 In oder to obtain a meaningful limit along the "base",
one has to show it is possible to get a sequence of consective non-trivial
limit connecting $x_- $ and $x_+$. The key point to prove this is to observe
that one can have isoperemetric inequality and monetonicity lemma for each ${\tilde u}_i$ projecting to $\xi$ in a small neighbourhood
of each point of ${\tilde M}$ as in the usual symplectic case. 
Now since each ${\tilde u}_i$ connects 
$x_-$ and $x_+$, and approaches to some of closed orbits along the  ends of 
the "base", 
its image projecting to $\xi $ is not very small. This  implies that
the non-trivial limit of above local convergence can be obtained. The 
of the analogy argument in symplectic case is used to produce 
intermediate bubbles, which can be
 found in [L?]. We refer the readers to the detail there there, which can be
easily
adapted here.

To do the step(iii), we define the potential "connecting bubble" 
$C_{i, R}= \Sigma_{i}\setminus
D_{i, R}\cup B_{i, R}$ for fixed $R$. Each componet  $C^k_{i, R}$ of 
$ C_{i, R} $ is a
sphere with several small discs removed and cylinder attached,
and connects the components of $
D_{i, R}$ and $B_{i, R}$. We may asssume that $\lim_{R\mapsto \infty}
\lim_{i\mapsto
\infty} E_{\lambda}({\tilde u }_{i}|_{C_{i}})\not= 0$. Then we get those intermediate 
connecting bubbles by local convergence of ${\tilde u}_i|_{C^k_{i, R}} $
with $R\rightarrow \infty.$
As mentioned above 
isoperemetric inequality and monotonicity lemma 
for  ${\tilde u}_i$ projecting to $\xi$ can be used to produce
non-trivial connecting bubbles. 

After this is done, we have $\lim_{R\mapsto \infty}\lim_{i\mapsto
\infty} E_{\lambda}({\tilde u }_{i}|_{ T_i})= 0$, 
where $T_{i}=\Sigma\setminus (B_{i,R}\cup
D_{i, R}\cup C_{i})$, i.e. there is no $E_{\lambda}$-energy loss
any more. We have got the full limit of the sequence ${\tilde u}_{i}$
along the compact direction. This is the projection of the sequence
to the contact manifold $M$ is already weakly convergent to the projection
of the limit map so far obtained. 

To get the full limit along the non-compact ${\ bf R}$-direction, we observe
that since there is no $E_{\lambda}$-energy loss anymore, given any
two of ends of any of above three parts, if presumely they should
joint together in the domain acording the above convergence scheme, but they
apporach to two closed orbits which lie on different ends of the target
( maybe in the different component of the target also), then the 
two closed orbits are the same,  and we get  trivial connecting map between
them ( maybe passing through several components of the target)
as part of the limit. Note that  only in the case there is already 
 some non-trivail
component lying on some component of the target, we may have to introduce
this kind of trivial connecting maps in the component in order to get a 
connected stable map. Therefore, the limit map so obtained is really a
 stable map defined before.

Finally, we note that in the next section we will prove that
%Since $\lim_{R\mapsto \infty}\lim_{i\mapsto
%\infty} E_{\lambda}({\tilde u }_{i}|_{ T_i})= 0$, 
when $R$ and $i$ large 
enough, each component $T^k_i$ of $T_{i}$, whose domain is equivalent to
$[-R^k_i, +R^k_i]\times S^1$, is exponetially close to the trivial ${\tilde
J}$-holomrphic map coming from some closed orbit $x$ when $T_{i}^k$
approaches to $x$.
\QED 
 
$\bullet$ {\bf Virtual co-dimension of the boundary of ${\overline {\cal M}}
(x_-, x_+; J)$}:

\begin{theorem}
The virtual co-dimension of the boundary components of 
${\overline {\cal M}}
(x_-, x_+; J)$
is at least one.
 In fact, the co-dimension of the stratum
of broken connecting ${\tilde J}$-holomorphic maps of two elements is one , 
and
co-dimension of any stratum whose elements contain
 bubble component is at least two.
\end{theorem}

{\bf proof}

The proof of this theorem depends on the index formula, which will
be proved in [L3].

Let $[{\tilde u}]$ be a typical element in the stratum.
It is sufficient to consider the follow two cases:

(i) The domain $\Sigma$ of $u$ is $\Sigma_{1}\cup \Sigma_{2}$ joint
together at one of the ends of $\Sigma_{1}$ and $\Sigma_{2}$.
 Each $\Sigma_{i}, i=1,2$
is $S^2$ with two marked points $-\infty$ and $+\infty$ treated as ends,
and we identify  
$\Sigma_{i}\setminus \{ \mbox end\}$ with $S^1\times {\bf  R}$ to give
two marked lines on $\Sigma$. The target $U$ of ${\tilde u}$ is
$ {\tilde M_{1}}\cup {\tilde M_{2}}$ joint at one of their ends.
${\tilde u}_{1}$ connects a closed orbit $x_{-, 1}$ on ${\tilde M}_{- ,1}$
and another closed orbit $x$ on ${\tilde M}_{+,1}={\tilde M}_{-, 2}$,
and ${\tilde u}_{2}$ connects the closed orbit $x$  
and another closed orbit $x_{+, 2}$ on ${\tilde M}_{+,2}.$

Note that $x\not = x_{-, 1}\not = x_{+, 2},$ and 
$x\not = x_{+, 2}.$ There are five dimensional
symmetries for each element $[{\tilde u}]$ is the above stratum, two
dimension coming from the ${\bf R}^1$-translations on each factor of the 
target and two dimensional ${\bf R}^1$-translations on each factor of the 
domain together with an $S^1$-action on the domain. We will
 slice out the $S^1$-action first. Let 
${\tilde {\cal M}}(x_-, \{x\}, \{x_+\}; J)$ be the moduli space of
marametrized broken connecting ${\tilde J}$-maps of two elements
as above. But we fix a parametrized $x_{-}$ and alow $x$ and $x_{+}$
vary
in their equivalent classes. The dimension of the
 symmetry group of the moduli space
is 4. It follows from the index formula in [L3] that the dimension of 
${\tilde {\cal M}}(x_-, \{x\}, \{x_+\}; J)$  is  same as the dimension of
${\tilde {\cal M}}(x_-, \{x_+\}; J)$  plus one, due to the one
dimensional 
possible choices of the element $x\in \{x\}$. Now a direct dimension
counting on the symmetries shows that in this case the codimension the
the boundary component of ${\overline {\cal M}}
(x_-, x_+; J)$ is one.

(ii) The second case  corresponds to the case that there is only
only one bubbling as described
in Lemma 3.4. There are two different subcases: (1) both of the principal
components are non-trivial; (2) the "new" principal components is trivial.
In the case (1), along the princial component, as parametrized map,
 there are three different 
possible parametrized closed orbits as asympotic limit along ends, but
is the case (2), there are only two of such closed orbits. On the other
hand, the dimension of the symmetry group of the two components lying in
the "left" component of the target ( not counting the $S^1$-action)
is 6  in the case (1) and 5 in the case (2).
Note that in the case (1) there are two connected components in
the "new' components of the target, while in the case (2) there is
only one according to our convention introduced before.
Again index formula in [L3] together with a direct  dimension
counting argument gives the desire conclusion in this case.

\QED

\section{Exponential Decay Estimate}

We have proved a version of compactness theorem for the moduli
space of stable ${\tilde J}$-holomorphic maps in last section. The result 
is not quite completed for its own ppurpose as well as for 
later applications. As we have shown before that a sequence of ${\tilde
J}$-holomorphic maps may develop bubbles and split into broken
connecting ${\tilde J}$-holomorphic maps. Unlike the usual Gromov-Floer
theory, these bubbles always have unremovable singularities. We showed
before that along the ends of singularities, the bubbles approach to
some closed orbits. For the purpose of moduli cycles in [L1], it is
important to know the rate of the ${\tilde J}$-holomorphic maps
approach to closed orbits either along their ends or along the ends of
the singularities of the bubbles. One of the main results of this
section is to prove that the rate of the approximation is exponential.
When $\dim M=3,$ this is proved by Hofer, Wysocki and Zehnder  in
 [HZW]. When $M$ has an $S^1$-symmetry, this is proved by
Li-Ruan  in [LiR]. We remark that the extra assumption of [LiR]
considerably simplified the analysis here. On the other hand, the
general case, even in $\dim M=3$, the argument in [HZW] is quite
involved. It turns out that the method of [HZW], suitably modified, can
be extended to the general case. We will carry out this generalization
somewhere else. In this section, we will give a more abstract and a
simpler proof.

To motivate the second main result of this section, we
note that one of the important ingredients of the proof of compactness
of the moduli space in the usual Gromov-Floer theory is an explicit
description about the behavior of the "connecting neck" near bubble
point. 

In our case, it is necessary to know that the behavior of the
"connecting necks" near the "connecting" closed orbit when a family of
${\tilde J}$-holomorphic maps develop, say, a bubble approaching to the
"connecting" closed orbit, or split into a broken ${\tilde J}$-
holomorphic maps of two elements joints at the closed orbit. More
precisely, if 
$${\tilde v}_i={\tilde u}_i|_{[-l_i, l_i]\times S^1}: [-
l_i, l_i]\times S^1\rightarrow {\tilde M}=M\times {\bf R}$$
 is the "neck"
part of ${\tilde u}_i$ such that the $M$-projection $v_i$ is close to the
closed orbit $x(t)$ with $c=\int_{S^1}x^*\lambda.$ We claim that
${\tilde v}_i$ is essentially the same as the trivial map $(s,
t)\rightarrow(x(t), cs)\in M\times {\bf R}$ restricted to $[-l_i,
l_i]\times S^1$.  In particular, the length of ${\bf R}^1$-
projective of ${\tilde v}_i$ differs from $2c\cdot l_i$ by at most a fixed
small constant.  Note that when $i\rightarrow\infty,
l_i\rightarrow\infty. $ 
 It turns out that this statement plays an important  role in
the compactness theorem. Recall that we have required that in the
definition of stable map, there is no unstable trivial connecting maps appeared
as components.  The justification of this 
 is based on the above statement.

\medskip

Let $x(t)$ be a closed orbit. ${\tilde u}=(u, a), {\tilde w}=(w, b)$ are
two ${\tilde J}$-holomorphic connecting maps: ${\bf R}^1\times
S^1\rightarrow {\tilde M}$ such that $\lim_{s\rightarrow +\infty}u(s,
t)=x(t)=\lim_{s\rightarrow -\infty}w(s, t). $ Assume that
$\lim_{s\rightarrow -\infty}u(s, t)=x_-(t) $ and $\lim_{s\rightarrow
+\infty}w(s, t)=x_+(t). $ Let ${\tilde v}^*_i=(v^*_i, f^*_i):{\bf
R}^1\times S^1\rightarrow {\tilde M}$ be a sequence of ${\tilde J}$-
holomorphic maps connecting $x_-(t)$ and $x_+(t)$ and locally convergent
to ${\tilde u}\cup{\tilde w}$.
Hence, $\lim_{s\rightarrow +\infty}u(s, t)=\lim_{s\rightarrow
-\infty}w(s, t)=x(t)$ of some closed orbit. 
 Note that the target ${\tilde M}$ of
${\tilde u}$ and ${\tilde w}$ should be thought as two different spaces
joint together at their ends. We will only prove our results for this
particular case. It is easy to see that the corresponding results for
the case that ${\tilde v}^*$ produces only one bubble can be proved in
an exactly the same way and the result for the general case can be
obtained by a simple combination of these two cases. 

The assumption that ${\tilde v}^*_i$ is locally $C^\infty$-convergent to
${\tilde u}\cup{\tilde w}$ implies that there exist $n_{i, j}\in {\bf
R}$, $m_{i, j}\in {\bf R}$, $j=1, 2$ such that ${\tilde v}^*_n(s+n_{i,
1}, t)+(0, m_{i, 1})$ is $C^\infty$-convergent to ${\tilde u}(s, t)$ and
$v^*_i(s+n_{i, 2}, t)=(0, m_{i, 2})$ is $C^\infty$-convergent to
${\tilde v}(s, t)$ for any compact subset of ${\bf R}^1\times S^1$. 

Now both $\{{\tilde u}(s+n, t)\}^\infty_{n=0}$ and $\{{\tilde v}(s-n,
t)\}^\infty_{n=0}$ are locally $C^\infty$-convergent to the trivial 
${\tilde J}$-holomorphic map
$(s, t)\rightarrow(x(t), cs)$, after translations in ${\tilde M}$. 
We conclude that $\exists N$ such that for any given $\epsilon >0$, when
$s>N$, $|D^\alpha\{u(s, t)-x(t)\}|<\epsilon=\epsilon_\alpha$ and $S<-N$,
$|D^\alpha\{w(s, t)-x(t)\}|<\epsilon=\epsilon_\alpha$ for any
$|\alpha|\geq 0$, and that $|D^\alpha\{a(s, t)-
c\dot s\}|<\epsilon=\epsilon_\alpha$, $|D^\alpha\{b(s, t)-c\dot
s\}|<\epsilon=\epsilon_\alpha$ for any $|\alpha|\geq 1.$ 

We now define ${\tilde v}_i(s, t)={\tilde v}^*_i(s+\frac{n_{i, 1}+n_{i,
2}}{2}, t).$ by the assumption on local convergence of ${\tilde v}^*_i$,
$n_{i, 1}\rightarrow -\infty$ and $n_{i, 2}\rightarrow +\infty$. Let
$l_i=\frac{1}{2}\{(n_{i, 2}-n_{i, 1}-2N\}.$ Then $l_i\rightarrow
+\infty.$ Then $v_i(-l_i, t)=v^*_i(N+n_{i, 1}, t)\rightarrow u(N, t)$
and $v_i(l_t, t)=v^*_i(-N+n_{i, 2}, t)\rightarrow w(-N, t).$

\begin{lemma}

When $i$ is large enough, for any $s\in (-l_i, l_i)$, $|D^\alpha\{v_i(s,
t)-x(t)\}|<2\epsilon$, $|\alpha|\geq 0$ and $|D^\alpha\{f_i(s, t)-
cs\}|<2\epsilon, $ $|\alpha|\geq 1. $ 
\end{lemma}

{\bf proof}

Since the proof of the two statements are similar, we will only prove
the first one. Assume that the first statement is not true. then there
exists  a sequence $(s_i, t_i)\in (-l_i, l_i)\times S^1$,
$i\rightarrow\infty$, such that $|D^\alpha\{v_i(s_i, t_i)-
x(t_i)\}|>2\epsilon.$ If $|s_i -(-l_i)|$ or $|s_i-l_i|$ are bounded, say 
$|s_i -(-l_i)|$ is bounded, then $v_i(s_i+s, t), s\in (-\delta, \delta)$
is $C^\infty$-convergent to $u({\underline N}+s, t)$ for some
${\underline N}>N$ and $s\in (-\delta, \delta)$, which implies that 
$$|D^\alpha\{v_i(s_i, t)-x(t)\}|<2\epsilon$$ when $i$ is large enough.
This is a contradiction. Hence we may assume that both $|s_i-(-l_i)|$
and $|s_i-l_i|\rightarrow\infty.$

Then ${\tilde v}_i(s_i+s, t)$ is still $C^\infty$-convergent for any
$(s, t)\in [- R,  R]\times S^1$, with fixed $R$. Let $R\rightarrow
\infty$ and patch all the local limit together, we get a ${\tilde J}$-
holomorphic map ${\tilde v}_\infty:{\bf R}^1\times S^1\rightarrow
{\tilde M}$ with $E_\lambda({\tilde v}_\infty)=0.$ This implies that
$v_\infty(s, t)=x(t).$ Therefore, $|D^\alpha(v_i(s_i, t)-
x(t))|<\epsilon$ when $i$ large enough. This is a contradiction again. 

\QED

To state one of our main results, we define 
$${\tilde v}_{i,+}(s, t)=(v_i(s-l_i, t), f_i(s-l_i, t)-f(-l_i, 0)+a(N,
0))$$ and 
$${\tilde v}_{i,-}(s, t)=(v_i(-s+l_i, t), f_i(-s+l_i, t)-f(l_i, 0)+b(-N,
0)).$$
Then ${\tilde v}_{i,+}(0,0)\rightarrow (u(N, 0), a(N, 0)),$ and
${\tilde v}_{i,-}(0,0)\rightarrow(w(-N, 0), b(-N, 0)).$

\vspace{3mm}

\noindent{$\bullet$ \bf Local Coordinate near x(t)}:

The $\lambda$-period of $x(t)$ is $\int_{S^1} x^*\lambda dt
=c$.  We have $\frac{dx}{dt}=c\dot X_{\lambda}(\alpha(t)). $ 
By rescaling the parameter $(s, t)$, we may assume that $c=1$.
 Let $\tau$ be the
minimal period of $x(t)$, i.e. $\tau >0$ is the minimal number such that
$x(t+\tau)=x(t).$ Under this assumption, 
given any point $z=x(t), t\in [0,\tau)$, we assign its
$\theta$-coordinate $\theta=\theta (z)=t.$ For simplicity, we will
assume further that $\tau=1.$ Hence 
$\theta\in
S^1={\bf R}/{\bf Z}$, and $x(\theta)=x(t), \theta\in S^1$ is the
simple closed orbit. Choose a global basis $\{e_1, \cdots, e_{2n}\}$ for
the symplectic bundle $(\xi, d\lambda)|_{x(\theta)}$ such that the map 
$$y=\Sigma y_i e_i(x(\theta))\in \xi\rightarrow (\theta, y_1, \cdots,
y_{2n})\in (S^1\times {\bf R}^{2n}, \omega_0)$$ 
gives rise a isomorphism between  the two trivial symplectic bundles
$(\xi, d\lambda)$ and $(S^1\times{\bf R}^{2n}, \omega_0)$ over $S^1.$
The local coordinate of $M$ near $x(\theta)$ is define by $(y, \theta)
\rightarrow exp_{x(\theta)}\Sigma y_ie_i$, where $y=(y_1, \cdots,
y_{2n})\in{\bf R}^{2n}, $ $\theta\in S^1.$ The exponential map is taken
with respect to the Riemanian metric ${ g}_{\tilde J}$. Note that we
may assume that $J|_{\xi|_{S^1}}$ corresponds to $J_0$ under above
iomorphism of the two symplectic bundles over $S^1=\{x(\theta)\}$.

Let $U$ be a small tube neighborhood of $x$ in $M$. With the above
coordinate $(y, \theta)$, then at any point $z\in U$,
$$T_z M={\bf R}\{\frac{\partial}{\partial \theta}\}\oplus {\bf
R}\{\frac{\partial}{\partial y_1}, \cdots, \frac{\partial}{\partial
y_{2n}}\}={\bf R}X_\lambda\oplus\xi_z.$$ 
Since at $y=0$, $\xi|_{y=0}={\bf R}\{\frac{\partial}{\partial y_1},
\cdots, \frac{\partial}{\partial y_{n}}\}_{y=0}$, the projection
$d\pi_y:T_zM\rightarrow {\bf R}\{\frac{\partial}{\partial y_1}, \cdots,
\frac{\partial}{\partial y_{2n}}\}_z$ when restricted to $\xi_z$, is an
isomorphism, when $|y|$ is small enough. Here $z=(y, \theta)$. We may
assume that any $z\in U$ has this property. Then we can find
$e_i=e_i(z)$ such that $d\pi_y(e_i)=\frac{\partial}{\partial y_i}$.
Since $d\pi_y(\frac{\partial}{\partial \theta})=0$,
$\frac{\partial}{\partial \theta}\not\in\xi_z$. Hence ${\bf
R}\{\frac{\partial}{\partial \theta}\}\oplus \xi_z=T_zM.$

For the application later, we need to compare  $e_i$ with
$\frac{\partial}{\partial y_i}$ and $X_\lambda$ with
$\frac{\partial}{\partial \theta}$. Let $e_i=\Sigma^{2n}_{i=1}\alpha_{i,
j}(z)\frac{\partial}{\partial y_i}+\alpha_{i, 0}(z)\frac{\partial}{\partial
\theta}, $ $X_\lambda=\Sigma^{2n}_{i=1}X_i(z)\frac{\partial}{\partial
y_i} +X_0(z)\frac{\partial}{\partial \theta}.$ Here $\alpha_{i, j}$ and
$X_i$ are functions defined on ${\bf R}^{2n}\times S^1=\{(y,\theta)\}.$
Fix $i$, since $e_i(0, \theta)=\frac{\partial}{\partial y_i}$, 

\begin{eqnarray*}
e_i(y, \theta)-\frac{\partial}{\partial y_i} & = & e_i(y, \theta) -
e_i(0, \theta)\\
& = & (\frac{\partial}{\partial \theta}, \frac{\partial}{\partial y_1},
\cdots, \frac{\partial}{\partial y_{2n}})\{\int ^1_0 \frac{d}{d\tau}
\left [\begin{array}{l}
\alpha_{i, 0}(\theta, \tau y)\\
\alpha_{i, 1}(\theta, \tau y)\\
\vdots\\
\alpha_{i, 2n}(\theta, \tau y)
\end{array}
\right ] d\tau\} \\
& = & (\frac{\partial}{\partial \theta}, \frac{\partial}{\partial y_1},
\cdots, \frac{\partial}{\partial y_{2n}})(\int d\alpha_i(\theta, \tau
y)d\tau)\left [\begin{array}{l}
y_1\\ 
\vdots\\
v_{2n}
\end{array}
\right  ].
\end{eqnarray*}

Here $d\alpha_i(\theta, y)=[\frac{\partial \alpha_{i, j}}{\partial
y_k}(\theta, y)]$ is the $(2n+1)\times (2n+1)$ matrix where the $(j,
k)$th element is $\frac{\partial\alpha_{i, j}}{\partial y_k}(\theta,
y).$ Similarly, 
$$X\lambda(y, \theta)-\frac{\partial}{\partial \theta}=
(\frac{\partial}{\partial\theta}, \frac{\partial}{\partial y_1}, \cdots,
\frac{\partial}{\partial y_{2n}})\int^1_0dx(\theta, \tau y)d\tau \left (
\begin{array}{l}
y_1\\
\vdots\\
y_{2n}
\end{array}
\right ),$$
where $dx(\theta, y)$ is the $(2n+1)\times (2n+1)$ matrix whose $(j, k)$
element is $\frac{\partial X_j}{\partial y_k}(\theta, y).$ Not that both
matrices $d\alpha_i$ and $dx$ has uniformly bounded norm for $(y,
\theta)\in U$. This proves

\begin{lemma}

For any $(y, \theta)\in U$, $\exists $ constant $C$ such that 
$$|e_i(y,\theta)-\frac{\partial}{\partial y_i}|<C\dot |y|, \quad
|X_\lambda(y,\theta)-\frac{\partial}{\partial \theta}|<C\dot |y|.$$

\end{lemma}

In the $(y, \theta, a)$-coordinate for $U\times {\bf R}\subset {\tilde
M}$, we write ${\tilde u}(s, t)=(u(s, t), a(s, t))$ and $u(s, t)=(y_u(s,
t), \theta_u(s, t))$. If there is no confusion, we will simply ommit the
subscript $u$ in $y_u$ and $\theta_u$. Similarly, we write $w(s,
t)=(y_w(s, t), \theta_w(s, t))$ and $v_i(s, t)=(y_{v_i}(s, t))$,
$\theta_{v_i}(s, t)$ in the $(y, \theta)$-coordinate.

\begin{lemma}

Let $\pi=\pi_{\xi}:TM={\bf R}\{X_\lambda\}\oplus\xi\rightarrow\xi$ be
the projection. Given any $v\in TM$, if
$$\pi(v)=\Sigma^{2n}_{i=1}c_i\frac{\partial}{\partial
y_i}+c_0\frac{\partial}{\partial \theta}=\Sigma^{2n}_{i=1}d_ie_i,$$
then $c_i=d_i, i=1, \cdots, 2n.$ 
\end{lemma}

\noindent{\bf Proof}

$\pi(v)=v-\lambda(v)X_\lambda.$ Let 
$$X_\lambda=\Sigma^{2n}_{i=0}X_i\frac{\partial}{\partial
y_i}+X_0\frac{\partial}{\partial \theta}$$ and 
$$v=\Sigma^{2n}_{i=1}v_i\frac{\partial}{\partial
y_i}+v_0\frac{\partial}{\partial \theta}.$$ 

Then 
\begin{eqnarray*}
\pi(v) & =   & \Sigma^{2n}_{i=1}(v_i-\lambda(v)\cdot
X_i)\frac{\partial}{\partial y_i}+(v_0-
\lambda(v)X_0)\frac{\partial}{\partial \theta}\\
& = &\Sigma^{2n}_{i=1}(v_i-\lambda(v)\cdot X_i)e_i +(v_0-
\lambda(v)X_0)\frac{\partial}{\partial \theta}\\
 & + & \Sigma^{2n}_{i=1}(v_i-
\lambda (v)X_i\cdot (\frac{\partial}{\partial y_i}-e_i).
\end{eqnarray*}

Now since $\pi_y(\frac{\partial}{\partial y_i}-
e_i)=\frac{\partial}{\partial y_i}-\frac{\partial}{\partial y_i}=0$, 
$$\frac{\partial}{\partial y_i}-e_i\in \ker\pi_y={\bf
R}\{\frac{\partial}{\partial\theta}\}.$$
Therefore $$\Sigma^{2n}_{i=1}(v_i-\lambda(v)X_i)(\frac{\partial}{\partial
y_i}-e_i)\in {\bf R}\{\frac{\partial}{\partial\theta}\}$$ and
$$\pi(v)=\Sigma^{2n}_{i=1}(v_i-\lambda(v)X_i)e_i,  \mbox{\,\,mod} ({\bf
R}\{\frac{\partial}{\partial\theta}\}).$$
But $\pi(v), e_{i}\in \xi$ and $\frac{\partial}{\partial\theta}\not\in \xi.$ This
implies that $\pi(v)=\Sigma^{2n}_{i=1}(v_i-\lambda(v)X_i)e_i.$
\QED

\vspace{3mm}

\noindent{$\bullet$ \bf Equation in the local coordinate}:

We only write the equation for ${\tilde u}$. Same expression is also
applicable to ${\tilde w}$ and ${\tilde v}_i.$ That ${\tilde u}$ is ${\tilde J}$-
holomorphic is equivalent to:
$$
\left \{
\begin{array}{lllr}
a_s & = &\lambda(u_t) & \quad \quad\quad\quad(a)\\
a_t & = & -\lambda(u_s) & \quad \quad\quad\quad(b)\\
\pi(u)\circ du\circ i & = & J(u)\pi(u) \circ du & \quad \quad\quad\quad(c)
\end{array}
\right.
$$
Let $M(y, \theta)$ be the $2n\times 2n$ matrix for the $d\lambda$-
compatible almost complex structure $J(y, \theta)$ with respect ot the
basis $\{e_1, \cdots, e_{2n}\}$. We will assume that $M(y, \theta)=J_0$,
the standard constant complex structure on ${\bf R}^{2n}$. That is
$J_0(e_i)=e_{i+n}$ and $J_0(e_{i+n})=-e_i, $ $1\leq i \leq n.$ As pointed
out in [HZW], the proof of the statements below for general $M$ can be reduced
to this case. For our purpose of this paper, we can even assume that
this is really true as we can make choice of $J$. The eqation (c) is
equivalent to $\pi(u_s)+J(u)\pi(u_t)=0.$ In local coordinate we have
\begin{eqnarray*}
\pi(u_s) & = &\Sigma^{2n}_{i=1}\{(y_i)_s-\lambda(u_s)X_i\}e_i\\
\pi(u_t) & = &\Sigma^{2n}_{i=1}\{(y_i)_y-\lambda(u_t)X_i\}e_i.
\end{eqnarray*}

Hence, 
$$
(y_s-\lambda(u_s)Y)+M(y_t-\lambda(u_t)Y)=0.
$$

Equivalently,

$$y_s+My_t+(a_t-a_s\cdot M)\cdot Y=0.$$

Here $y=\left [\begin{array}{l}
y_1\\
\vdots\\
y_{2n}
\end{array}
\right ]$ and $Y=\left [
\begin{array}{l}
X_1\\
\vdots\\
X_{2n}
\end{array}
\right ],  $
and $M=J_0$.

\vspace{2mm}

We have shown that  
$$Y(y, \theta)=\{\int^1_0 dY(\tau y, \theta)d\tau\}\left
(\begin{array}{l}
y_1\\
\vdots\\
y_{2n}
\end{array}
\right )$$ and $dY(y, \theta)$ is the $2n\times 2n$ matrix whose $(j,
k)$-element is $\frac{\partial X_j}{\partial y_k}.$ 

Denote $\int^1_0dY(\tau y, \theta)d\tau$ by $DY(y, \theta).$ Then
$$y_s+My_t+\{(a_t-a_sM)\cdot DY\}\cdot y =0.$$

Denote $\{a_t-a_sM\}\cdot DY(y(s, t), \theta(s, t))$ by $S(s, t).$ 
We define $S_\infty =-J_0\cdot dY(0, t).$

\begin{lemma}

When $s> N$, $|S(s, t)-S_\infty(s, t)|<C\cdot\epsilon$ and $|S_s(s,
t)|<C\cdot\epsilon $ for the given $\epsilon$ and some constant $C.$
Same conclusion for $w$ and $v_i$ when $s<-N$ or $s\in (-l_i, l_i)$
respectively.
 
\end{lemma}

\noindent{\bf Proof}:

We only prove the  statement for $u$. 

When $s>N$, 
\begin{eqnarray*}
|D_s\{u(s, t)-x(t)\}|& = &|D_s\{(y(s, t), \theta(s, t))-(0, t)\}|\\
& = & |D_s(y(s, t), \theta(s, t))|<\epsilon.
\end{eqnarray*}

Note that in the $(y, \theta)$-coordinate, $x(t)=(0, t)$ since $c=1.$
Similarly, when $s>N$, 
$$|D_s\{a_t(s, t)-\frac{\partial}{\partial
t}(cs)\}|=|D_sa_t(s, t)|<\epsilon$$ and 
$$|D_s\{a_s(s, t)-\frac{\partial}{\partial s}(cs)\}|=|D_sa_s(s,
t)|<\epsilon.$$ 
This implies that $|D_sS(s, t)|<C\cdot \epsilon$ for some constant $C$
depending only on $||DY(y, 0)||_{C^1}$ on $U.$

When $s>N$, $|a_t(s, t)|=|D_t(a(s, t)-cs)|<\epsilon$ with $c=1$ and 
$|a_s(s, t)-1|=|D_s(a(s,t)-cs)|<\epsilon, $ we have 
$$|(y(s, t), \theta(s, t))-(0, t)|<\epsilon.$$ This implies that 
$$|S(s, t)-\{-J_0DY(0, t)\}|<\epsilon.$$
But 
\begin{eqnarray*}
-J_0DY(0, t) & = & -J_0\int^1_o dY(0, t)d\tau\\
& = & -J_0 dY(0, t)=S_\infty(t).
\end{eqnarray*}
\QED

\begin{lemma} 

$S_\infty(t)$ is a $2n\times2n$ symmetric metric and all the eigenvalues
of the self-adjoint elliptic operator
$A_\infty:L^2_1(S^1, {\bf R}^{2n})\rightarrow L^2(S^1, {\bf R}^{2n})$
 defined by $A_\infty: z\rightarrow -
J_0 \frac {dz}{dt}-S_\infty \cdot z$, are non-zero. 

\end{lemma}

\noindent{\bf Proof}:

Let $\Psi_t$ be the flow of $X_\lambda.$ Hence
$$
\left\{
\begin{array}{lr}
\frac{d\Psi_t(z)}{dt}=X_\lambda(\Psi_t(x)) & \quad\quad\quad (*)\\
\Psi_0(z)=z, \forall z\in M. &
\end{array}
\right .
$$

If $z_0=(0, 0)$ in $(y, \theta)$-coordinate then $\Psi_t(z_0)=
\Psi_t(0, 0)=(0, t)=x(t).$ Hence $z_0=\Psi_1(z_0)$ is a fixed point of
$\Psi_1$. Note that the flow $\Psi_t$ preserves the decomposition
$TM={\bf R}\{X_\lambda\}\oplus\xi$, and that along $x(t)=(0, t)$,
$X_\lambda=\frac{\partial}{\partial \theta}$ and $\xi={\bf
R}\{\frac{\partial}{\partial y_1}, \cdots, \frac{\partial}{\partial
y_{2n}}\}. $ Differentiating  eqaution (*) above, we get 
$$\frac{dD\Psi_t}{dt}=DX_\lambda(\Psi_t)\circ D\Psi_t. \quad \quad (**)$$

Now given $v, w\in \xi_{(0, 0)}\subset T_{(0,0)}M$, since $\Psi_t$
preserves $d\lambda =\omega$, We have 
\begin{eqnarray*}
\omega (J(\Psi_t)_*(v), J(\Psi_t)_*(w_0)) & =& \\
\omega((\Psi_t)_*(v), (\Psi_t)_*(w)) & =& \omega(v, w).
\end{eqnarray*}

Differentiating this, we get
$$\omega(J(\frac{d}{dt}D\Psi_t)(v), JD\Psi_t(w)) + \omega(JD\Psi_t(v),
J(\frac{d}{dt}D\Psi_t)(w))=0.$$

Here we used that $J$ is constant along $\xi|_{x(t)}.$ Use equation (**),
we get 
$$\omega(JDX_\lambda(\Psi_t)\circ D\Psi_t(v), JD\Psi_t(w))+
\omega(JD\Psi_t(v), JDX_\lambda(\Psi_t)\circ D\Psi_t(v))=0.$$
Let $v_t=D\Psi_t(v), w_t=D\Psi_t(w).$ Then 
$${g}_J(JDX_\lambda(\Psi_t)(v_t), w_t)={ g}_{J}(v_t, 
JDX_\lambda(\Psi_t)(w_t)).$$
Let $t=1$, then $\Psi_t(z)=z$ for any $z=(0, \theta).$ It is easy to see
that 
$$DX_\lambda(0, t)=\left (
\begin{array}{ll}
dY(0, t) & 0\\
0 & 1
\end{array}
\right ),$$ 
$$JDX_\lambda(0, 0)(v_1)=JdY(0, 0)(v_1),$$ and
$$JDX_\lambda(0,0)(w_1)=JdY(0,0)(w_1).$$ This implies that 
$S_\infty=-J_0 dY(0,0)$ is symmetric. Then general case can be proved by
a coordinate change on $t.$ Therefore, $A_\infty=-J_0\frac{d}{dt}-
S_\infty:L^2_1(S^1, {\bf R}^{2n})\rightarrow L^2(S^1, {\bf R}^n)$ is a
self-adjoint elliptic operator. We want to show that $0$ is not an
eigenvalue of $A_\infty.$ Given $0\not = z\in L^2_1(S^1, {\bf R}^{2n})$, 
$A_\infty(z)=0$ is equivalent to 
$$\frac{dz}{dt}=J_{0}S_\infty(t)z=dY(0, 1)z, \quad\quad\quad\quad (***)$$
with   $z(t+1)=z(t).$ As before let $z_0=(0,0).$ Then $\Psi_t(z_0)=(0,t)$ in
$(y,\theta)$-coordinate. We write
$$D\Psi_t(z_0)=\left (
\begin{array}{ll}
R(t) & 1\\
0 & 1
\end{array}
\right 
)$$
with respect to the basis $$\{\frac{\partial}{\partial y_1},
\cdots,\frac{\partial}{\partial y_{2n}},
\frac{\partial}{\partial\theta}\}.$$ 
The equation (**) implies
$$\frac{dR(t)}{dt}=dY(0,t)\cdot R(t).\quad\quad\quad\quad (****).$$
If $w(t)\not=0 $ is a solution of (***), then $w(t+1)=w(t).$ 

Define ${\tilde w}(t)=R(t)\cdot w(0)$, then (****) implies
$$\frac{d{\tilde w}}{dt}=dY(0,t){\tilde w}(t).$$
Since ${\tilde w}(0)=R(0)w(0)=w(0), $ we have ${\tilde w}(1)=w(1)=w(0)
$, 
i.e. $w(0)$ is an eigenvalue of $R(1)$ with eigenvalue $1$. This implies
that $d\Psi_1(z_0)$ has an eigenvector of eigenvalue $1$ along
$\xi_{z_0}.$ Conversely, if $v$ is an eigenvector of $R(1)$ with
eigenvalue $1$, then $w(t)=R(t)\cdot v$ solves (***) with 
$w(1)=R(1)\cdot v=v=R(0)\cdot v=w(0).$ Therefore, we get an eigenvector
of $A_\infty$ of eigenvalue $0$. 

It follows from this and previous lemma that 

\begin{lemma}

There exists a constant $\delta>0$, such that when $N$ and $i$ large
enough, for ${\tilde u}$ and ${\tilde w}$, with $s>N$ or $s<-N$
respectively, 
$$\|(-J_0\frac{d}{dt}-S(s,t))\cdot z\|\geq 2\delta\|z\|, \forall z\in
L^2_1(S^1, {\bf R}^{2n}).$$ 
For ${\tilde v}_i$ with $s\in (-l_i, l_i)$, same conclusion holds. 
\end{lemma}

We denote $-J_0\frac{d}{dt}-S(s, t)$ by $A(s):L^2_1(S^1; {\bf
R}^{2n})\rightarrow L^2(S^1; {\bf R}^{2n}).$ Note that $|A(s)-
A^*(s)|=|S-S^*|\leq |S-S_\infty|+|S^*-S^*_\infty|<c\cdot \epsilon,$ when
$s>N$ or $s<-N$ for ${\tilde u}$ or ${\tilde w}$, or $S\in (-l_i, l_i)$
for ${\tilde v}_i$. 

We now establish the exponential decay estimate for the $y$-components of
${\tilde u}$, ${\tilde w}$ and ${\tilde v}_i$. We will use $y=y(s)=y(s,
-)\in L^2_1(S^1; {\bf R}^{2n})$ to denote the $y$-components of ${\tilde
u}$, ${\tilde w}$ or ${\tilde v}_i$. Let $g(s)=\frac{1}{2}<y(s), y(s)>.$

\begin{lemma}

When $N$ and $i$ large enough, for $s>N$ or $s<-N$ for ${\tilde u}$ or
${\tilde w}$, and for $s\in (-l_i, l_i)$ for ${\tilde v}_i$, we have
$$g^{''}(s)\geq \delta^2 g(s).$$

\end{lemma}

\noindent{\bf Proof}:

\begin{eqnarray*}
g'(s) &= &<y'(s), y(s)>.\\
g''(s) & = & <y_s, y_s>+<(y_s)', y(s)>\\
& = & <A\cdot y, A\cdot y>+<\frac{\partial}{\partial s}(-
J_0\frac{dy}{dt}-S\cdot y), y(s)>\\
& = & <A\cdot y, A\cdot y>+<y_s, A^*y>-<S_sy, y>\\
& = & 2\|A\cdot y\|^2+<A\cdot y, (A^*-A)\cdot y>-<S_sy, y>\\
& \geq & 2\|Ay\|^2-C\epsilon \|Ay\|\cdot \|y\|-C\epsilon\|y\|^2\\
& =& \|Ay\|(2\|Ay\|-C\epsilon\|y\|)-C\epsilon\|y\|^2\\
& \geq & \delta\|y\|^2(2\delta-C\epsilon-\frac{C\epsilon}{\delta})\\
& \geq & \delta^2\|y\|^2=\delta^2g(s).
\end{eqnarray*}

Here we use the fact that $C$ and $\delta$ are uniformly bounded for all
$s$ and $\epsilon$ can be made as small as possible by the suitable
choice of $s$ in the lemma. 

\QED

For ${\tilde u}$ and ${\tilde w}$, since $s\in [N, +\infty)$ or $s\in (-
\infty, -N]$, and $g(s)\rightarrow 0$ as $s\rightarrow\pm\infty$, the
above lemma together with the usual elliptic estimate applied to each 
$[s_i, s_{i}+1]\times S^1$ implies that 

\begin{lemma}

The $y$-component $y(s, t)$ of ${\tilde u}$ satisfies:
$$\|y(s)\|^2_{L^2}\leq \|y(N)\|^2_{L^2}\cdot e^{-\delta(s-N)}, \quad
s>N.$$
Moreover, there exists a constant $C=C_\alpha,$ with $ |\alpha|\geq 0$, such
that $$|D^\alpha y(s, t)|<C_\alpha\cdot e^{-\delta(s-N)}.$$

Similar conclusion holds for ${\tilde w}$. 
\end{lemma}

To get corresponding estimate for ${\tilde v}_i$, we note that since
$\lim_{s\rightarrow +\infty} y_u(s, t)=0=\lim y_w(s, t),$
we may assume that $\|y_u(N)\|_{L^2}=\|y_w(-N)\|_{L^2}.$ This implies
that $\|y_{v_i}(-l_i)\|_{L^2}$ is very close to
$\|y_{v_i}(l_i)\|_{L^2}$, when $i$ large enough. For simplicity, we may
assume that $c_+=g(l_i)=\|y_{v_i}(l_i)\|^2=\|y_{v_i}(-l_i)\|^2=g(-
l_i)=c_-.$
Let $c=c_+=c_-$, and denote $l_i$ by $l$. Define $h(s)=a\cdot(e^{-
\delta s}+e^{\delta s}), $ with $a=\frac{c}{e^{-\delta l}+e^{\delta
l}}.$ Then $h(-l)=h(l)=c$, and $h''(s)=\delta^2 h(s). $

Define $f=g-h$. Then $f''(s)\geq \delta^2\cdot f(s),$ for $ s\in (-l, l)$ and
$f(-l)=f(l)=0.$ The maximal principle implies that $f(s)\leq 0, s\in (-l
,l).$ Hence $g(s)\leq\frac{c\cdot (e^{-\delta s}+e^{\delta s})}{e^{-
\delta l}+e^{\delta l}}.$ 

Now define $g_+(s)=g(s-l)$, and $g_-(s)=g(l-s)$, $s\in (0, l).$ Then 
\begin{eqnarray*}
g_+(s) & \leq & c\cdot\frac{e^{-\delta (s-l)}+e^{\delta (s-l)}}{e^{-
\delta l}+e^{\delta l}}\\
& \leq & 2\cdot c\frac{e^{-\delta s}\cdot e^{\delta l}}{e^{\delta
l}}=2g_+(0)\cdot e^{-\delta s}\\
 & = & 2c_+ e^{-\delta s}, \quad \quad\quad\quad s\in[0,l ].
\end{eqnarray*}

Similarly, $g_s(s)\leq 2c_- e^{-\delta s}$. 

Note that since $c_+$, $c_-$ are close to $\|y_u(N)\|^2_{L^2}$ and
$\|y_w(-N)\|^2_{L^2}$ which are fixed, we get exponential decay of
$g_+(s)$ and $g_-(s)$. For the general case when $c_+\not = c_-$, we have
$g_+(s)\leq 2(c_++c_-+\epsilon)\cdot e^{-\delta s}$ for some fixed small
$\epsilon$ when $i$ large enough. 

Define ${\tilde v}_{i, +}={\tilde v}_i(s-l_i, t)$ and ${\tilde v}_{i, -
}(s, t)={\tilde v}_i(l_i-s, t)$, and let $y_+$, $y_-$ be the
corresponding $y$-components.  We have

\begin{lemma}

When $i$ large enough, 
$$|y_{\pm}(s, t)|^2_{L^2}\leq 2(\|y_u(N)\|^2_{L^2}+\|y_w(-
N)\|^2_{L^2}+\epsilon)\cdot e^{-\delta s}, \quad s\in (0, l_i).$$
Moreover, $\exists C=C_\alpha, $ $|\alpha |\geq 0$ such that 
$$|D^\alpha y_{\pm}(s, t)|<C\cdot e^{-\delta s}, \quad \quad s\in (0,
l_i).$$
\end{lemma}

We now study the behavior of the $(a, \theta)$-component of ${\tilde u},
{\tilde w}$ and ${\tilde v}_i$. 

We have shown before that when $s>N$ or $s<-N$, for $u$ and $w$, and
$s\in(-l_i, l_i)$ for $v_i$, $|D(u(s, t)-x(t))|<\epsilon.$ Since $|Dy(s,
t)|<\epsilon$, this implies that $|\partial_t\theta -
1|=|\partial_t\theta-\partial_t x(t)|<\epsilon$ and
$|\partial_s\theta|=|\partial_s\theta-\partial_s(x(t))|<\epsilon.$ Let
${\cal P}:U\subset{M}\rightarrow {\bf R}^1\times S^1=\{(a,
\theta)\}$ be the projection of the $(a, y,\theta)$-coordinate chart 
$U$ to  $(a, \theta)$-coordinate chart ${\bf R}^1\times S^1$ given by 
$(a, y, \theta)\rightarrow (a, \theta).$ Then ${\underline u}={\cal P}\circ
{\tilde  u},
$ ${\underline w}={\cal P}\circ {\tilde w} $ and ${\underline v}_i={\cal P}\circ {\tilde v}_i$ are
local diffeomorphisms from $[-N, +\infty)\times S^1$, $(-\infty, -
N]\times S^1$ and $(-l_i, l_i)\times S^1$ to ${\bf R}^1\times S^1.$
Since 
$$|\partial_s a(s, t)-1|=|\partial_s (a(s, t)-cs)|<\epsilon$$
$|\partial_t a(s, t)|<\epsilon$ for these  values of $s$ in the above range
\begin{eqnarray*}
|a(s, t)-a(s_0, t_0)| &\geq & |a(s, t)-a(s_0, t)|-|a(s_0, t)-a(s_0, t_0)|\\
& \geq & \frac{1}{2}|s-s_0|-\epsilon.
\end{eqnarray*}

This implies that   ${\underline u}$, ${\underline w}$ and ${\underline v}_i$ 
are proper.. Hence they are covering maps from open cylinders $(N,
+\infty)\times S^1, $ $(-\infty, -N)$ or $(-l_i, l_i)\times S^1$ to
their images in 
${\bf R}\times S^1.$ Assume that the degree of the covering is $m$. 

Let $\pi_m:{\bf R}\times S^1\rightarrow {\bf R}\times S^1$ be the
standard $m$-fold covering induced from the corresponding covering of
$S^1$ to $S^1$. Write  ${\underline u}$, ${\underline w}$ and ${\underline v}_i$ 
as $(a, \theta).$ We will study ${\underline v}_i$ first. We will only
derive the equation for ${\underline v}_i=(a, \theta).$ The same formula
is also applicable for ${\underline u}$ and ${\underline w}$. 

Let $q_0={\underline v}_i(-l_i, 0)=(a_0, \theta_0)$ and 
${\tilde q}_0=(a_0, {\tilde \theta}_0)\in \pi_m^{-1}(q_0)$
with ${\tilde \theta}_0\in [0, 1)$ being the smallest of such ${\tilde
\theta}_0.$ Define ${\underline V}_i $ to be the unique lifting of 
${\underline v}_i$ sending $(-l_i, 0)$ to ${\tilde q}_0$. We drop the 
subscript of ${\underline V}_i$ from now on. Then ${\underline V}:(-l_i,
l_i)\times S^1\rightarrow {\bf R}\times S^1$ is an embedding. Note that
the length of the image of $a$-projection of ${\underline
V}(\{l_i\}\times S^1)$ and ${\underline V}(\{l_i\})\times S^1)$ is less
than $\epsilon. $ Hence the image of ${\underline V}$ in ${\bf R}\times
S^1$ is almost a standard cylinder of the form $[0, L]\times S^1. $ We
want to prove that $|L-2l_i|$ is uniformly bounded and tends to zero
when $i$ and $N$ tends to infinity. Since $\pi_m$ preserves $a$-length,
this also implies the corresponding statement for ${\underline v}_i.$ 

To this end, define the complex structure ${\underline i}={\underline
i}(s, t)$ on the image of ${\underline V}$ by the identification:
$${\underline i}(s, t) = dV(s, t)\circ i \circ \{d{\underline V}(s,
t)\}^{-1}:T_{{\underline
V}(s, t)}({\bf R}^1\times S^1)\rightarrow T_{{\underline
V}(s, t)}({\bf R}^1\times S^1). $$
 Then ${\underline V}$ is $(i, {\underline i})$-holomorphic, i.e.
$$d{\underline V}\circ i={\underline i}({\underline V})\cdot d{\underline
V}.$$
Equivalently, $$\frac{\partial {\underline V}}{\partial s}+
{\underline i}(s, t)\frac{\partial {\underline V}}{\partial t}
=\frac{\partial {\underline V}}{\partial s}+{\underline i}({\underline
V})\frac{\partial {\underline V}}{\partial t}=0.$$
Switch to $v_{i, +}$ or $v_{i,-}$ and consider the corresponding
${\underline v}_{i, +}, {\underline v}_{i, -}$ and ${\underline V}_+,
{\underline V}_-$ and associated ${\underline i}(s, t). $ By abusing our
notations, we will still use ${\underline i}(s, t)$ to denote the complex
structure in these cases. 

\begin{lemma}

For $s\in (-l_i, l_i)$, there exists a constant $C=C_\alpha$ independent
of $i$, such that $|D^\alpha({\underline i}(s, t)-i)|<C_\alpha\cdot e^{-
\delta s}.$

\end{lemma}

\noindent {\bf Proof}:

Since $\pi_m$ is a local diffeomorphism, if we define ${\underline
I}={\underline I}(s, t):T_{{\underline  v}(s,t)}({\bf R}^1\times S^1)\rightarrow
T_{{\underline v}(s, t)}({\bf R}^1\times S^1)$ by the formula:
$d{\underline v}(s, t)\circ i\circ \{d{\underline v}(s, t)\}^{-1}$, then 
${\underline i}(s, t)=d{\pi}_m^{-1}\circ {\underline I}(s, t)\circ
d\pi_m$. Therefore, we only need to prove the corresponding statement
for ${\underline I}(s, t)$. Now ${\underline v}$ is $(i, I)$-
holomorphic, i.e. $d{\underline v}(s, t)\circ i=I(s, t)\circ d{\underline v}(s, t).$
In terms of the basis $\frac{\partial {\underline v}}{\partial s}(s,t)
$, $\frac{\partial {\underline v}}{\partial t}(s, t), $ 

$$I(s,t)(\frac{\partial {\underline v}}{\partial s}, 
\frac{\partial{\underline v}}{\partial
t})=(\frac{\partial {\underline v }}{\partial s}, \frac{\partial
{\underline v}}{\partial t})\cdot
\left (
\begin{array}{lr}
0 & -1\\
1 & 0
\end{array}
\right )
.$$

We need to find the expressions for $\frac{\partial{\underline v}}
{\partial s}$ and $\frac{\partial {\underline v}}{\partial t}$ in 
terms of $(\frac{\partial}{\partial a}, \frac{\partial}{\partial \theta}).$

\vspace{3mm}

\noindent{\bf Sublemma}

$$(\frac{\partial {\underline v}}{\partial s}, \frac{\partial
 {\underline v}}{\partial t})=(\frac{\partial }{\partial a}, 
\frac{\partial}{\partial \theta})\cdot
\left \{
\left (
\begin{array}{lr}
a_s & -a_t\\
a_t & a_s
\end{array}
\right )
+ O(e^{-\delta s})
\right \}.$$

\noindent {\bf Proof:}

\begin{eqnarray*}
\frac{\partial {\underline v}}{\partial s} & = & d{\cal P}\circ d{\tilde
v}_i (\frac{\partial }{\partial s})\\
& = & d{\cal P}(a_s\frac{\partial}{\partial a}+\{(v_i)_s-\lambda(v_i)_s
X_\lambda\} +\lambda(v_i)_sX_\lambda).
\end{eqnarray*}

Let $(v_i)_s-\lambda(v_i)X_\lambda(v_i)_s=\Sigma^{2n}_{k=1}
c_k\frac{\partial}{\partial y_k}+c_0\frac{\partial}{\partial \theta}.$
Then 
\begin{eqnarray*}
(v_i)_s-\lambda(v_i)X_\lambda(v_i)_s & =& \Sigma^{2n}_{k=1} c_k e_k\\
& = &\Sigma^{2n}_{k=1} c_k\frac{\partial}{\partial y_k}
+\Sigma^{2n}_{k=1} c_k\cdot (e_k-\frac{\partial }{\partial y_k}).
\end{eqnarray*}

Now $c_k$ is uniformly bounded and $|e_k(u(s, t)-
\frac{\partial}{\partial y_k})|<C\cdot |y(s, t)|<C\cdot e^{-\delta s}.$
Similarly, 
$$\lambda(v_i)_s X_\lambda=\lambda(v_i)_s(X_\lambda -\frac{\partial}{\partial
\theta})+\lambda(v_i)\cdot\frac{\partial}{\partial \theta},$$
and $$|X_\lambda(u(s, t))-\frac{\partial}{\partial \theta})|< c\cdot
|y(s, t)|<c\cdot e^{-\delta s}.$$
This implies that 
$$\left \{
\begin{array}{l}
\frac{\partial {\underline v}}{\partial s}=a_s\frac{\partial}{\partial
a}+\lambda\{(v_i)_s\}\frac{\partial }{\partial \theta}+O(e^{-\delta
s})\\
\frac{\partial {\underline v}}{\partial t}=a_t\frac{\partial}{\partial
a}+\lambda\{(v_i)_t\}\frac{\partial}{\partial \theta}+O(e^{-\delta s}).
\end{array}
\right .
$$

Now $\lambda\{(v_i)_s\}=-a_t$ and $\lambda(v_i)_t=a_s.$ The conclusion
follows. 
\QED

Let
$$A=\left (
\begin{array}{lr}
a_s & -a_t\\
a_t & a_s
\end{array}
\right ), \mbox{ and } O=O(e^{-\delta s}).$$ 

Then in terms of the basis $(\frac{\partial }{\partial s},
\frac{\partial}{\partial t})$:
$${\underline I}(s, t)=(A+O)\cdot J_0(A+O)^{-1}.$$

Since $AJ_0=J_0A$, ${\underline I}(s, t)=J_0+O(e^{-\delta s})$. This
proves the lemma for $\alpha =0.$ The general case with $|\alpha|\geq 1$
can be proved similarly. 
\QED

Still work with ${\tilde v}_{i, +}$ and the corresponding ${\underline
V}$. Now ${\underline V}:(0, l_i)\times S^1\rightarrow {\bf R}^1\times
S^1.$ By ${\bf R}$-translation, we may assume that ${\underline
V}(0,0)=(0, \theta_0).$ Note that $\theta_0\rightarrow 0$ when
$i\rightarrow \infty.$ Consider the unique lifting of ${\underline V}$
from the universal covering $(0, l_i)\times {\bf R}^1$ of $(0,
l_i)\times S^1 $ to the universal covering ${\bf R}^1\times {\bf R}^1 $
of ${\bf R}^1\times S^1, $ which sends $(0,0)$ to $(0, \theta_0)$,
$0\leq \theta_0\leq 1.$ We still denote it by ${\underline V}.$ Then
$({\underline V}-Id):(0, l_i)\times {\bf R}^1\rightarrow {\bf R}^1\times
{\bf R}^1 $, and since both ${\underline V}$ and $Id$ commutes with deck
transformations induced by $\theta\rightarrow\theta+1, $, ${\underline
V} -Id$ is periodic on the second factor of $(0, l_i)\times {\bf R}^1$
of period 1. Let $\Phi={\underline V}-Id:(0, l_i)\times S^1\rightarrow
{\bf R}^2.$ Then 
$$\frac{\partial\Phi}{\partial s}=\frac{\partial{\underline V}}{\partial
s}-\frac{\partial (Id)}{\partial s}=-\{i+O(e^{-s})\}\frac{\partial
{\underline V}}{\partial t}-i\frac{\partial (Id)}{\partial t}.$$
Since $\frac{\partial {\underline V}}{\partial t}$ is bounded, we have 
$$\frac{\partial \Phi}{\partial s}+i\frac{\partial \Phi}{\partial
t}+O(e^{-\delta s})=0.$$

\begin{pro}

$|\Phi(s, t)|<C $ for all $s\in (0, l_i)$, where $C$ is bounded by the
initial value of $ O(e^{-\delta s})$,  $|\Phi(0)|$ and
$|\frac{\partial}{\partial t}\Phi|$. All of them tend to zero as $i$ and
$N$ tends to infinity.
\end{pro}

\noindent{\bf Proof:}

Let $\phi(s)=\int_{S^1}\Phi(s, t)dt.$ Then 
$$\frac{d\phi}{ds}=-\int_{S^1}O(e^{-\delta s})dt = f(s) (=O(e^{-\delta
s})).$$
Hence $\phi(s)=\phi(0)+\int^s_0 f(\tau)d\tau.$ If $|f(s)|<d\cdot 
e^{-\delta s}$, $s\in (0, l_i)$, then $|\int^s_0
f(\tau)d\tau|<\frac{d}{\delta}.$ 

Now let $\Psi(s,t)=\Phi(s,t)-\phi(s).$ Then 
$$\int_{S^1}\Psi(s,t)dt=\phi(s)-\phi(s)=0.$$ 
Let $C_1=\max |\frac{\partial}{\partial t}\Phi(s,t)|=\max |\frac
{\partial}{\partial t}\Psi(s,t)|$. Clearly $|\Psi(s,t)|<2C_1.$ Hence 
$$|\Phi(s,t)|<|\Psi(\theta)(s, t)|+ |\phi (s)|<|\phi(0)|+\frac{d}{\delta}+2C_1.$$

\QED

We reamrk that this proposition is the precise statement we mentioned before
on the behavior of the "connecting neck" along the non-compact $a$-
direction, which is used  in the previous section to justify why 
 it is possible  to  get  the compactification of the
moduli space without introducing the  unstable trivial connecting maps.  
 
For ${\tilde u}$ and ${\tilde w}$, we get more. We only prove the result
for ${\tilde u}$. Define ${\underline u}$, ${\underline U}$ and
${\Phi}={\underline U}-Id:(N, -\infty)\times S^1\rightarrow
{\bf R}^2$ as above. We have
$$\frac{\partial \Phi}{\partial s}+i\frac{\partial \Phi}{\partial
t}+O(e^{-\delta (s-N)})=0.$$

Let $O(e^{-\delta(s-N)}=f(s,t).$ We identify the image ${\bf R}^2$ of
$\Phi$ and $f$ with ${\bf C}.$ Then the standard complex structure 
$i=\left (
\begin{array}{lr}
0 & -1\\
1 & 0
\end{array}
\right )$ on ${\bf R}^2 $ is identified with the multiplication by
imaginary number $i.$ Fix $s$, let
\begin{eqnarray*}
\Phi(s,t ) & = & \Sigma_{n\in Z}\phi_n(s)e^{int}\\
f(s,t) & = & \Sigma_{n\in Z} f_n(s) e^{int}
\end{eqnarray*}
be the Fourier expansion
of $\Phi(s, -)$ and $f(s, -)$. Then $\phi'_n-n\phi_n+f_n=0,  n\in Z.$
Note that $|f_n(s)|<C\cdot e^{-\delta (s-N)}$. In particular, when $n=0,
$ $\phi'_0(0)=f_0(s).$ Hence
\begin{eqnarray*}
\phi_0(s) & = & \phi_0(N)+\int^s_Nf_0\\
& = & \phi_0(N)+\int^\infty_Nf_0 -\int^\infty_s f_0\\
& = & s_0-\int ^\infty_sf_0,
\end{eqnarray*}
where $s_0=\phi_0(N)+\int^\infty_Nf_0$ is a constant. Now
$$|\int ^\infty_sf_0(s)ds|<C\cdot\int^\infty_s e^{-\delta(\tau-
N)}d\tau<\frac{C}{\delta}e^{-\delta (s-N)}.$$
Hence $|\phi_0(s)-s_0|<C_1\cdot e^{-\delta (s-N)}$. Now let
$\Psi(s,t)=\Phi(s,t)-\phi_0(s)$ and $r(s,t)=f(s,t)-f_0(s).$ Then
$$\frac{\partial \Psi}{\partial s }+i\frac{\partial \Psi}{\partial
t}+r(s,t)=0.$$

Note 
$$<i\frac{\partial\Psi}{\partial t}, i\frac{\partial \Psi}{\partial
t}>=<\Sigma_{n\not =0}n\phi_n(s)\cdot e^{int}, \Sigma_{n\not
=0}n\phi_n(s)\cdot e^{ins}>\geq \Sigma_{n\not
=0}\phi^2_n(s)=<\Psi,\Psi>.$$

Define $g(s)=\frac{1}{2}<\Psi(s),\Psi(s)>.$ Then 
\begin{eqnarray*}
g''(s) & = & <\Psi'(s),
\Psi'(s)> +<\Psi''(s), \Psi(s)>\\
& = & 2<i\Psi_t, i\Psi_t>+<r,r>+<i\Psi_t, r>+<r, i\Psi_t>-<r_s, \Psi>\\
& \geq & 2\|\Psi\|(\|\Psi\|-\|r\|-\|r_s\|).
\end{eqnarray*}

Note that both $|r(s,t)|$ and $|r_s(s,t)|\leq C\cdot e^{-\delta(s-N)}$.
Let $h(s)=\|r(s,t)\|+\|r_s(s,t)\|\leq 2C\cdot e^{-\delta(s-N)}.$ Then,
if $\|\Psi\|(s)> 2h(s)$, we have 
$$g'(s)\geq 2\|\Psi\|(\|\Psi\|-\frac{1}{2}\|\Psi\|)\geq g(s).$$ Now the set
$P=\{s \quad | \quad \|\Psi\|(s)> 2h(s)\}$ is open and is a countable
union of $(s_i, s_{i+1}) $ such that $\|\Psi\|(s_i)=2h(s_i)$ and
$\|\Psi\|(s_{i+1})=2h(s_{i+1}).$ Assume that $\delta >1$, then the
argument before to prove Lemma \ref{???} implies that, for $s\in (s_i,
s_{i+1})$, 
$g(s)\leq 4\cdot C\cdot e^{-\delta (s-N)}.$ On the other hand, if
$s\not\in P$, then $\|\Psi\|(s)\leq 2h(s)=4 C\cdot e^{-\delta (s-
N)}.$ 

We conclude that $g(s)\leq C_1\cdot e^{-\delta (s-N)}.$ As before,
applying elliptic estimate to get higher order estimate, we get

\begin{pro}

Let ${\tilde u}(s, t)=(u(s,t), a(s,t))$ be a ${\tilde J}$-holomorphic
map such that $\lim_{s\rightarrow\infty}u(s,t)=x(t)$ of a closed orbit
of $\lambda$-period $c$. Let $u(s,t)=(y(s,t),\theta(s,t))$ in the local
$(y,\theta)$-coordinate near $x(t).$ Then there exist positive
constants $N, C=C_\alpha $  and $\delta$ such that 
\begin{eqnarray*}
|D^\alpha y(s,t)| & < &C_\alpha\cdot e^{-\delta s},\\
|D^\alpha\{(a(s,t),\theta(s,t))-(cs+d_1,ct+d_2)\}| & < & C_\alpha\cdot
e^{-\delta s}
\end{eqnarray*}
for some suitable constants $d_1\in {\bf R}$ and $d_2\in (0,\tau)$,  where
$\tau$ is the minimal period of $x(t).$
\end{pro}

\section{Some possible applications}

The following are some immediate possible applications. Here we will only
briefly indicate the reasons for these applications and refer the reader
to the forth coming papers on each of these topics.

$\bullet$ {\bf (A) Index homology in contact geometry}:

We have already outlined the index homology in contact geometry by
using the moduli space of connecting pseudo-holomorphic maps.
Note that the moduli space of connecting maps used here has an one
dimensional symmetry of $S^1$- rotations. At the same time, their
asymptotic ends of closed orbits also have the $S^1$-symmetry.
It is possible to remove the symmetry by using the connecting
${\tilde J}={\tilde J}_{t}, t\in S^1$ holomorphic maps with $t$ dependent
${\tilde J}$.  This will lead to a special  Bott-type index homology,
even the contact structure is generic.

$\bullet$ {\bf (B) Additive  quantum homology in contact geometry}:

The index homology we defined is an analogy of the usual Floer
homology in symplectic geometry. We now outline 
a quantum homology in contact geometry by a different way to use 
the moduli space.

Let $M_{a}\subset {\tilde M}= M\times {\bf R}^1$ be the section 
$M\times \{a\}$ in ${\tilde M}$. Given any singular chain ${\alpha}$
in $M=M_{a}$ consider the moduli space 
${\tilde {\cal M}}(x_-, x_{+}; \alpha, M_{a})/ S^1$, which  is a subset of 
${\tilde  {\cal M}}(x_-, x_{+})/S^1$ whose element $u$ satisfies the
condition that $ u(z_{0})\in \alpha (\in M_{a})$. Now consider another marked
 point $z_{1}$
in $u$  lying on a fixed marked line 
and define the obvious evaluation map 
$e_{\alpha}=e_{\alpha; a, b, z_{1}}: \cup_{x_{-}, x_{+}}
 {\tilde {\cal M}}(x_-, x_{+}; \alpha, M_{a})/ S^1 \rightarrow M_{b}$.
Note that each element $u$ with the two marked points without any 
non-compact symmetry anymore.

The intuition here is that by letting $a$ being very negative, and
$b$ very positive, we  flow the singular chain $\alpha$ lying almost
in the negative end to get a collection of singular chains  almost
in the
positive end.

We now define additive quantum homology by defining the chain
complex generated by singular chains $\alpha$ in $M$ with the boundary
$D(\alpha) ={\partial} (\alpha) \pm e_{\alpha}$, where ${\partial} (\alpha)$
is just the usual boundary map of singular homology.
By the property of the moduli space established in this paper and 
[L1], we have $D^2=0$. One can show that the homology so defined
is independent of the choices involved and is an invariant
of $M$.

$\bullet$ {\bf (C) Gromov-Witten invariants in contact geometry
and ring structure in  the index cohomology }:

Whence the index homology is  defined, we can define G-W invariant
in exact the same fashion as the G-W invariant in the usual
quantum homology and Floer homology.
Namely given closed orbits $x_{-}=(x_{1,-}, \cdots, x_{k,-})$ and $
x_{+}=(x_{1,+}, \cdots, x_{l,+})$, we define G-W invariant 
$\Psi_{k, l} (x_{-}, x_{+})$ by counting $J$-homomorphic map $u$
in ${\tilde M}$  from the domain $S^2$ with $k+l$ punctures
such that along $k$ negative ends $u$ approaches to $x_{-}$ and
along $l$ positive  ends $u$ approaches to $x_{+}.$ 
 
As in the usual GW-invariant in quantum and  Floer homology, one can show that
the invariant so defined at chain level descends to the 
homology.

By using the invariant $\Psi_{2,1}$, one can define a ring structure
in the index cohomology, 
which can be thought as a quantum product
for the contact manifold.

One can also extend to definition of G-W invariants by introducing
another set of marked points $z=(z_{1}, \cdots, z_{n}), z_{i}\in S^2$,
and require that $u(z_{i})\in C_{i}$ of some prescribed cycles in 
${\tilde M}.$

Using the special case of three marked point  invariants with only one $z,$
we get an action of $H_{*}(M)$ on the index homology, i.e. the index
homology is a module over $H_{*}(M)$.

There are obvious generalization of theses constructions, such as 
higher genus G-W invariants, coupling with gravity and so on.

Note that the product structure should be thought as an essential
part of the structure of these index homologies as there are many cases
where the additive index homology is infinitely generated.

$\bullet$ {\bf (D) Relative quantum  homology}:

Give a compact symplectic manifold $(P, \omega)$ with a contact boundary,
 let the boundary
be $M$ with the compatible contact structure $\lambda=i_{X} \omega$, where
$X$ is the  contact field i.e. ${\cal L}_{X}\omega = \omega .$ We now
glue ${\tilde M}$ to $P$ along the boundary. By using a suitable choice
of $\phi$ mentioned before, we get a new symplectic manifold with
a cylindrical end.

The chain complex of the quantum  homology of a symplectic
manifold $P$ with contact boundary $M$ is generated by the
  pair $(\alpha, \beta)$  where $\alpha $ is a singular chain in $P$
and $\beta$ is a singular chain in ${\tilde M}$. The boundary operator
$D=(D_{1},D_{2}).$ Here $D_{2}(\beta)$ is defined same as the one in (A)
above. $D_{1} ({\alpha})=\partial (\alpha) \pm e_{\alpha}$.
The definition of $  e_{\alpha}$ here is also similar to the one in (B).
But
we use the moduli space of $J$-holomorphic maps, the domain of whose
elements is ${\bf C}$ being treated as a half sphere with an half
infinite cylinder attached, to flow the singular chain $\alpha $ in $P$
to get a collection of singular chains $e_{\alpha}$ in $M$. Again
we have $D^2=0$. Now there is an obvious embedding of the chain
complex of the quantum homology of $M$ defined in (B) to the chain
complex we just defined. We define the chain complex of the relative
quantum homology as the quotient of this pair.

Note that unlike (B) above, in the case that the contact boundary $M$ 
of $P$ is  concave, we may not be able to get a desired uniform
energy bound. In this case we need some extra assumption such as
$\omega$ is exact.
 
 Note that there are some obvious algebraic constructions related to
these chain complexes, such as the induced  long exact sequences related
these three homologies  and Mayer-Vietoris sequence of these homologies.
More general, assume that we  can decompose a compact symplectic
manifold in sequence of increasing symplectic  sub-manifolds with (convex)
contact type boundaries, we can associate the sequence a filtration
of chain complexes defined above. Then there is a associated spectral
sequence associated to the filtration. 

It is an interesting question to study further these algebriac
constructions to incorperate the multiplicative structures and to study
their relation to the quantum homology of a symplectic manifold. 
It seems that this will give
a new way to compute quantum homology of a symplectic manifold.

%Floer  homology for this new non-compact symplectic manifold by using those
%Hamiltonian functions, whose behavior along the end is same as the ones 
%described in (A).

$\bullet$ {\bf (E) Bott-type index homology, $S^1$-invariant contact
manifold and Weinstein conjecture}:

 We have assumed so far  that the contact form $\lambda$ is generic so that
the set of closed orbits is discrete. We can relax this condition by only
requiring that $\lambda$ is of Bott-type. Then the set of closed orbits
decomposes into an union of different components, each being a manifold.
Note that the period of any element in a component is the same by Stokes
theorem. In the symplectic case, in this situation, Ruan and Tian
developed a Bott-type Floer homology.  One can develop a  similar
construction in this case. As remarked in $A$, we have two different versions
 of the Bott-type
homology.

One of our motivation to consider Bott-type index homology is
to answer the question that if the index  homology so defined is always trivial.
%We mentioned in (A) that the Hamiltonian function $H$ used there has no
%critical points. This
%leads to the question if the associated Floer homology is always trivial.

By using the Bott-type index homology, one can compute the index homology
when the contact manifold appears as a regular zero locus of 
a local Hamiltonian function on some symplectic manifold, which generates
a local $S^1$-action.

For simplicity, let  $(P, \omega)$ be a compact symplectic manifold
with a $S^1$ Hamiltonian action generated by a Hamiltonian function $H$. 
Assume that $a$ is a regular value of $H$. Let $P^a= H^{-1}(a)$ and
$P_{a}=P^{a}/ S^1$. Under some assumption,
$P^{a}$ is a contact manifold whose contact structure is specified by 
$\omega$.  
In fact  the contact structure on $P^a$ can be chosen to be 
$S^1$-invariant.
We define a $S^1$-invariant contact form as follows. Note that
$P^{(a-\epsilon, a+\epsilon)}$ is a $S^1$ bundle over 
$P_{(a-\epsilon, a+\epsilon)}$, where 
$P^{(a-\epsilon, a+\epsilon)}= H^{-1} ((a-\epsilon, a+\epsilon))$
and $P_{(a-\epsilon, a+\epsilon)}=H^{-1} ((a-\epsilon, a+\epsilon))/S^1.$
Chose a connection.  We can lift any vector field $X$, which
is transversal to $P_a$ in $P_{(a-\epsilon, a+\epsilon)}$ to an $S^1$-equivariant
vector field ${\tilde X}$. We define the $S^1$-invariant contact form
$\lambda =i_{{\tilde X}}\omega$. By adjusting $X$, we may assume that
$\lambda(X_{H})=1$. That is $\lambda$ the connection $1$-form for 
the $S^1$-bundle. 
Hence, $\lambda$ is a contact form if the curvature
$d\lambda$ is  positive. Now the set of  closed orbits of the contact
manifold $(P^a, \lambda)$ of period $1$ is just $P_{a}$ and the images
of these closed orbits foliated $P^a$ itself. All other components of the set
of closed orbits are just  copies  of this one according to different
periods. We are in the situation of Bott-type index homology. The 
chain complex of Bott-type homology is generated by singular chains in 
some components of the set of closed orbits and the boundary map
is the combination of the usual boundary map for singular homology together
 with a "connecting" map by using the connecting  $J$-holomorphic maps
between two components of the set of closed orbits to flow the singular
 chain.  In our case, due to the extra $S^1$-symmetry in the moduli space
of $J$-connecting maps, the second part, the part of the "connecting"
map, of the boundary map has no contribution. Hence, the 
Bott-type index homology  is just infinitely many
copies of the usual homology of the symplectic
quotient $P_{a}$. In view of the invariance of Bott-type index homology,
this also compute the index homology for the contact structure. In particular,
we proved the non-vanishing of index homology in this case.
%Similarly, one can also compute Bott-type Floer homology and hence the
%Floer homology in the symplectization in this case.

As a corollary, we proved Weinstein conjecture for this case.

It would be interesting to study the relationship of 
the product structure in the contact
manifold $P^a$ with the quantum homology of its quotient, the
symplectic
manifold $P_{a}$.

$\bullet$ {\bf (F) Gluing formula for G-W invariants}:

Give a compact symplectic manifold $(P, \omega)$, assume that there is
a contact type hypersurface $M\subset P$ such that $M$ cuts $P$ into two
 pieces $P_{-}$ and $P_{+}$ with the common boundary $M$. As in (D), we can
glue ${\tilde M}$ to each of $P_{-}$ and $P_{+}$ to form two non-compact
symplectic manifolds $P^-$ and $P^+$ with cylindrical ends.
 As in [LR], we can prove a gluing formula for G-W invariants, which
relates the G-W invariants of $P$ with the G-W invariants in $P^+$,
$P^-$ and ${\tilde M}$.

The idea is the following:

One first collect all  $J$-holomorphic map $u$ in   $P^+$, $P^-$ or 
${\tilde M}$
with
the property that $u$ approaches to some of closed orbits lying on the ends
of $P^+$, $P^-$ or ${\tilde M}$ along its punctures, then select among
them those $u$ can be glued along those closed orbits.

Note that unlike in [LR], we do not require any local $S^1$ Hamiltonian 
action.

$\bullet$ {\bf (G) Low dimensional contact manifold}
A special feature of a three dimensional compact manifold is that
it always has a contact structure. Hence the index homology and 
additive quantum homology is well-defined associated to the contact
structure. It would be very interesting to investigate if the invariants
we defined here are actually topological invariants. There are various
 different forms of this type of questions. In view of the work of Taubes
on the relationship of the SW-invariants and GW-invariants, 
one  may hope to get similar results for contact 3-fold and symplectic
four manifold with contact type boundary. Our result should serve as one
of the basis to formulate this type of results.

\medskip

We make the following final remark. As we mentioned before, one of the main
results of this paper and [L3] is about the virtual co-dimension of the boundary
of the moduli space, which is the foundation of the applications outlined
in this section. This result is the consequence of the compactness theorem
proved in this paper and the index formula, which will be proved in [L3].
To obtain the result, the index formula we need here is  different
from the usual one appeared in Bott-type Floer homology due to the extra
dimension of the asymptotic $R^1$-motion of a connecting pseudo-holomorphic 
maps along
 the ends (closed orbits).

On the other hand, the main body of this paper, the proof of 
 the compactness theorem,  is independent of the desired index formula.
In fact, the new phenomenon appeared in the bubbling described in Lemma 3.4
and the Definition 4.1 on equivalence of stable maps concerning how to
count symmetries in target already opens the door for various possible
applications.

\ 


\begin{thebibliography}{W}   


\bibitem[EH]{eh}
Y. Eliashberg,
 Invariants in contact topology, 
{\it ICM 1998} {\bf Vol II} (1998), pp. ~327-338. 

\bibitem[FO]{fo}                        
Fukaya and Ono,
Arnold conjecture and Gromov-Witten invariants,
{\it Topology} (1999).
%



\bibitem[F]{f}
 A. Floer, Symplectic fixed points and holomorphic spheres, 
{\it Comm. Math. Phys. } {bf 120}(1989), pp. ~575-611. 

\bibitem[G]{g}
M. Gromov,
Pseudo holomorphic curves in symplectic manifolds,
{\it Invent. Math.} {\bf 82} (1985), pp. ~307-347.



\bibitem[H]{h}
H. Hofer, Pseudo holomorphic curves in symplectizations
with applications to Weinstein conjecture in dimension three.
{\it   Invent. Math. } {\bf 114} (1993), pp. ~515-563.
  

\bibitem[HWZ]{hwz}
H. Hofer, K. Wysocki, E. Zehnder, Holomorphic curves in symplectizations I:
Asympotics. {\it  Ann. I. H. P. Analyse Non Lineaire } {\bf 13} (1996), 
pp. ~337-379.
                      
\bibitem[LiR]{lir}
A. Li and Y. Ruan,
 Symplectic surgery and Gromov-Witten invariants of Calabi-Yau
3-folds I,
{\it Preprint} (1998). 

% 
 \bibitem[LiT]{lit}
J. Li and G. Tian,
Virtual moduli cycles and GW-invariants  of general symplectic manifolds,
{\it Proceedings of 1st IP conference at UC, Irvine} (1996).


 

\bibitem[L1]{l1}     %
G.Liu,                                     
Virtual Moduli cycles in the symplectization, 
 {\it In preperation.} 

%\bibitem[L2]{l2}     %
%G.Liu,    
%Index   homology and its multiplicative structures, 
% {\it In preperation.} 

%\bibitem[L3]{l3}     %
%G.Liu,    
%Additive quantum homology and  relative quantum homogy in contact geometry, 
% {\it In preperation.} 

%                                                            
%\bibitem[L4]{l4}     %
%G.Liu,    
%Bott-type index homology in contact manifold 
%and Weinstein conjecture, 
%{\it In preperation.} 

%\bibitem[L5]{l5}     %
%G.Liu,    
%Relative G-W invariant and its  gluing formula, 
%{\it In preperation.} 


\bibitem[L3]{l3}     %
G.Liu,  Fredholm theory  of the linearized ${{\bar \partial }}$-operator
and additivity of
 the index formula, 
{\it  Preprint.} 

\bibitem[LT]{lt}
G. Liu and G. Tian,
 Floer homology and Arnold conjecture,
{\it  JDG } {\bf 49}  (1998),pp. ~1-74.

 
 
\bibitem[RT]{rt}  
Y. Ruan and G. Tian, 
Bott-type   symplectic Floer cohomology and its multiplication structures,
{\it preprint } (1994).

 
%\bibitem[V]{v}
% C. Viterbo, Fuctors and computations in Floer homology with
%applications, I,
%{\it  GAFA, } {\bf Vol 9}  ( 1999),
%pp. ~985-1033.
% 
                 
\bibitem[T]{t}
C. Taubes, The Seiberg-Witten invariants and symplectic forms,
{\it  Math. Res. Letters} {\bf 1}  ( 1994) 
pp. ~809-822.




%
\end{thebibliography}
\end{document}